\documentclass[journal,10pt]{IEEEtran}
\IEEEoverridecommandlockouts

\usepackage[utf8]{inputenc}
\usepackage[T1]{fontenc} 
\usepackage[american]{babel}
\usepackage[flushleft]{caption}
\usepackage{graphicx}
\usepackage{listings}
\usepackage{float}
\usepackage{stfloats}
\usepackage{amsmath,amssymb,exscale}
\usepackage{blindtext, graphicx}
\usepackage{verbatim}
\usepackage{algorithm}
\usepackage{algorithmicx}
\usepackage{algpseudocode}
\usepackage{fancyvrb}
\usepackage{bera}
\usepackage{amsmath}
\usepackage{verbatim}
\usepackage{mathtools}
\usepackage{subcaption}
\usepackage{lipsum}
\usepackage{bm}
\usepackage{geometry}
\usepackage{color}
\usepackage{float}
\usepackage{lettrine}
\usepackage{pifont} 

\usepackage{times}
\newcommand{\revise}{\textcolor{black}}
\usepackage{cite}
\usepackage{scalerel}
\usepackage{tikz}
\usepackage[bookmarks=false]{hyperref}
\usetikzlibrary{svg.path}
\usepackage{authblk}
\renewcommand{\thefootnote}{\arabic{footnote}}

\title{Toward Dual-Functional LAWN: Control-Aware System Design for Aerodynamics-Aided UAV Formations}
  \author{
    \IEEEauthorblockN{Jun Wu, \IEEEmembership{Graduate Student Member, IEEE,}  Weijie Yuan, \IEEEmembership{Senior Member, IEEE}, Qingqing Cheng, \IEEEmembership{Member, IEEE}, \ and  Haijia Jin, \IEEEmembership{Graduate Student Member, IEEE}
  }
  \thanks{

  J. Wu, W. Yuan, and H. Jin are with the School of Automation and Intelligent Manufacturing, Southern University of Science and
  Technology, Shenzhen 518055, China (email: wuj2021@mail.sustech.edu.cn; yuanwj@sustech.edu.cn; jinhj2024@mail.sustech.edu.cn).

  Q. Cheng is with the School of Electrical Engineering and Robotics,
 Queensland University of Technology, Brisbane, QLD 4000, Australia (email: qingqing.cheng@qut.edu.au).

  }
  }
\begin{document}
\maketitle
\begin{abstract}
Integrated sensing and communication (ISAC) has emerged as a pivotal technology for advancing low-altitude wireless networks (LAWNs), serving as a critical enabler for next-generation communication systems. This paper investigates the system design for energy-saving unmanned aerial vehicle (UAV) formations in dual-functional LAWNs, where a ground base station (GBS) simultaneously wirelessly controls multiple UAV formations and performs sensing tasks. To enhance flight endurance, we exploit the aerodynamic upwash effects and propose a distributed energy-saving formation framework based on the adapt-then-combine (ATC) diffusion least mean square (LMS) algorithm. Specifically, each UAV updates the local position estimate by invoking the LMS algorithm, followed by refining it through cooperative information exchange with neighbors. This enables an optimized aerodynamic structure that minimizes the formation’s overall energy consumption. To ensure control stability and fairness, we formulate a maximum linear quadratic regulator (LQR) minimization problem, which is subject to both the available power budget and the required sensing beam pattern gain. To address this non-convex problem, we develop a two-step approach by first deriving a closed-form expression of LQR as a function of arbitrary beamformers. Subsequently, an efficient iterative algorithm that integrates successive convex approximation (SCA) and semidefinite relaxation (SDR) techniques is proposed to obtain a sub-optimal dual-functional beamforming solution. Extensive simulation results confirm that the `V'-shaped formation is the most energy-efficient configuration and demonstrate the superiority of our proposed design over benchmark schemes in improving control performance.

\end{abstract}
\section{Introduction}

The low-altitude wireless network (LAWN) has been anticipated to be a transformative paradigm for industry revolution in logistics, environmental monitoring, and smart agriculture. It capitalizes on low-altitude devices, particularly unmanned aerial vehicles (UAVs), aiming to offer superior flexibility, rapid deployment, and cost efficiency \cite{jiang20236g}. To unlock its full potential, LAWN requires seamless integration of real-time environmental sensing and reliable wireless communication, both of which are essential for ensuring the safe and intelligent operation of UAV-assisted networks. Specifically, effective sensing enables UAVs to perceive their surroundings, detect obstacles, and monitor environmental conditions, facilitating autonomous navigation and mission execution. Furthermore, reliable communication is indispensable for transmitting control commands and maintaining coordination among UAVs and ground infrastructure.

Integrated sensing and communication (ISAC) has emerged as a key enabler for LAWN, providing a spectrum-efficient framework that enables UAVs to simultaneously perform sensing and communication using shared spectral and hardware resources \cite{10418473,10791452,10098686,10663800}. By jointly optimizing these two functionalities, ISAC can effectively alleviate spectrum congestion, minimize hardware redundancy, and enhance decision-making efficiency in large-scale UAV networks. Given these inherent advantages, ISAC-LAWN has attracted growing research interest, with significant efforts dedicated to developing advanced beamforming techniques \cite{10681882}, UAV trajectory optimization \cite{10168298}, and spectrum-efficient resource allocation strategies \cite{10659350}. To name a few, the study in \cite{10158322} explored power control and path planning within a multi-UAV cooperative network, wherein UAVs not only track a designated target but also provide communication services to ground units, enhancing tracking accuracy and network throughput simultaneously. Moreover, the work in \cite{9916163} proposed a dual-functional beamforming design for a multi-antenna UAV-based ISAC system, investigating the fundamental trade-offs between sensing and communication in beam optimization. As a step forward, a recent solution in \cite{cheng2024networked} developed an ISAC-LAWN framework, where ground base stations (GBSs) work collaboratively to transmit signals to facilitate communication with multiple authorized UAVs while simultaneously detecting unauthorized objects. Despite these advancements, significant challenges remain in fully achieving LAWNs' capability. To be specific, LAWNs may suffer from environmental uncertainties in practical deployment, e.g., wind disturbance and weather impact, degrading both sensing and communication reliabilities. To address these variations, an effective control policy is essential to dynamically adjust the UAV's flight heading, speed, and attitude for performance improvement. 

Recognizing the above challenges, plenty of research efforts have been devoted to control design within UAV networks in recent years. In \cite{shima2009uav}, the authors proposed a cooperative decision-making and control framework to coordinate multiple UAVs in performing complex tasks. In particular, each UAV was assigned specific missions and feasible trajectories to optimize overall group performance. However, such mission-level control strategies, like waypoint navigation and power allocation, primarily operate at the \textit{outer loop} level, which governs high-level planning but does not directly ensure the stability of UAV attitude and flight dynamics, i.e., the \textit{inner loop} control \cite{10.1145/3301273}. To overcome this limitation, the study in \cite{meng2024communication} focused on an integrated closed-loop model that considers sensing, communication, and control while accounting for the coupling effects between uplink and downlink communications. An optimization framework was developed to jointly optimize the control policy and resource allocation for enhanced system efficiency. Additionally, the work in \cite{10050119} investigated a UAV-enabled sensing-communication-computing-control integrated network, in which multiple UAVs wirelessly control remote robots to guarantee system stability. To maintain effective control, the authors formulated a linear quadratic regulator (LQR) minimization problem to optimize system performance. \revise{Although significant advances have been made in the development of control strategies for system stabilization, most existing approaches offer only superficial treatment of the coupling among sensing, communication, and control, and fail to characterize the fundamental trade-offs inherent. This limitation undermines their relevance to LAWNs, where large-scale aerial deployments, dynamic sensing demands, and environmental uncertainty necessitate a unified framework that simultaneously integrates and optimizes all three functionalities.}

Additionally, the sustainability and reliability of LAWNs are constrained by limited onboard energy and computational resources. Therefore, energy-efficient design is paramount to ensure long-term viability and effectiveness in real-world applications. Conventionally, energy efficiency (EE) has been commonly adopted as a key metric to enhance the sustainability of UAV-based systems. For example, the study in \cite{8119562} explored a UAV-enabled data collection system, aiming to minimize the total energy consumption of all sensor nodes while ensuring reliable data acquisition from each sensor node. Note that, \cite{8119562} primarily optimized the transmission-related power, which is significantly lower than the energy required for UAV movement. Since UAV energy consumption is predominantly dictated by flight speed, focusing solely on transmission power may lead to suboptimal energy efficiency. Given the aforementioned, recent studies have explored joint optimization strategies considering both transmission and movement-related energy consumption, as elaborated in \cite{7888557,10158322,10159441,6965778}.\revise{
While existing approaches contribute to extending network lifetime, the optimization objectives are typically confined to minimizing the individual energy consumption of each UAV, without accounting for the potential benefits of cooperative flight dynamics in multi-UAV systems. In practical low-altitude deployments, UAVs inevitably interact with surrounding airflow patterns, which can be strategically exploited to enhance overall energy efficiency. For fixed-wing platforms, airflow over the wings induces wake vortices that generate an upwash region behind the wingtips \cite{HUMMEL1983321}. By flying within this region, a trailing UAV can experience reduced aerodynamic drag, leading to significant savings in propulsion energy \cite{weimerskirch2001energy}. This aerodynamic mechanism, well-documented in avian flocking behavior, offers a compelling opportunity for realizing energy-efficient cooperative formations in UAV networks.  However, research on aerodynamic-based cooperative energy-saving flights for LAWNs has not yet been reported in the existing literature.}

Motivated by the above, this paper investigates the trade-offs among sensing, communication, and control to optimize energy-efficient formation flight in  LAWNs. To achieve this, we propose a UAV formation approach that leverages aerodynamic principles to minimize flight energy consumption. Moreover, we consider a multi-antenna GBS that acts as both a radar and communication transmitter, enabling simultaneous wireless control of multiple UAV formations and target sensing in the low-altitude airspace via unified information and sensing waveforms. Our objective is to design the transmit beamforming to optimize the control performance while satisfying the predetermined radar sensing requirements. To the best of the authors' knowledge, this is the first attempt to leverage aerodynamics for designing control-oriented dual-functional beamformers tailored for UAV formation. The main contributions of this paper are summarized as follows:
\begin{itemize}
    \item 
    We develop a distributed UAV formation framework that effectively capitalizes on the aerodynamic upwash effect to reduce the induced air drag. By leveraging the adapt-then-combine (ATC) diffusion least mean square (LMS) approach, each UAV first estimates its position and iteratively refines its accuracy by exchanging information with neighboring UAVs, contributing to an energy-saving aerodynamic structure.
    \item To ensure stable UAV formation,  we derive the LQR cost as a control performance metric and formulate a maximum LQR minimization problem to guarantee control fairness across all formations, which is subject to power constraints and sensing requirements. To address this non-convex problem, we first develop a closed-form LQR expression for arbitrary beamforming. We then propose a computationally efficient algorithm by utilizing the successive convex approximation (SCA) and semidefinite relaxation (SDR) techniques to achieve a near-optimal beamforming solution.
    \item Extensive simulation results verify that the proposed design achieves superior LQR performance with lower energy consumption in UAV formation, compared to benchmarks. The results demonstrate that the `V' formation is the most energy-efficient flight configuration, leveraging aerodynamic upwash to enhance movement efficiency. Additionally, the results reveal the trade-offs between sensing accuracy and control performance, effectively capturing the intricate coupling between system control and wireless resource allocation.
    
\end{itemize}
The remainder of this paper is structured as follows: Sec. \ref{sec2} introduces the system model, and Sec. \ref{sec3} presents the developed energy-saving formation framework. The proposed dual-functional beamforming design is illustrated in Sec. \ref{sec4}, while Sec. \ref{simulation} presents the simulations to validate the effectiveness of the proposed approach. Finally, Sec. \ref{seccon} concludes this paper.

 \textit{Notations:} The boldface lowercase letter and boldface capital letter denote the vector and the matrix, respectively. The superscript $(\cdot)^{\mathrm{T}}$ and $ (\cdot)^{\mathrm{H}}$ stand for the transposition and Hermitian operations, respectively. We use $\mathcal{CN} (\mu,v)$ to depict a complex Gaussian distribution of mean $\mu$ and variance $v$.  $\mathbb{E(\cdot)}$ represents the
 statistical expectation. $|\cdot|$ and $|| \cdot||$ indicate the absolute value and Euclidean norm, respectively. 

\section{System Model} \label{sec2}
\begin{figure}[t]
    \centering
    \includegraphics[width=0.87\linewidth]{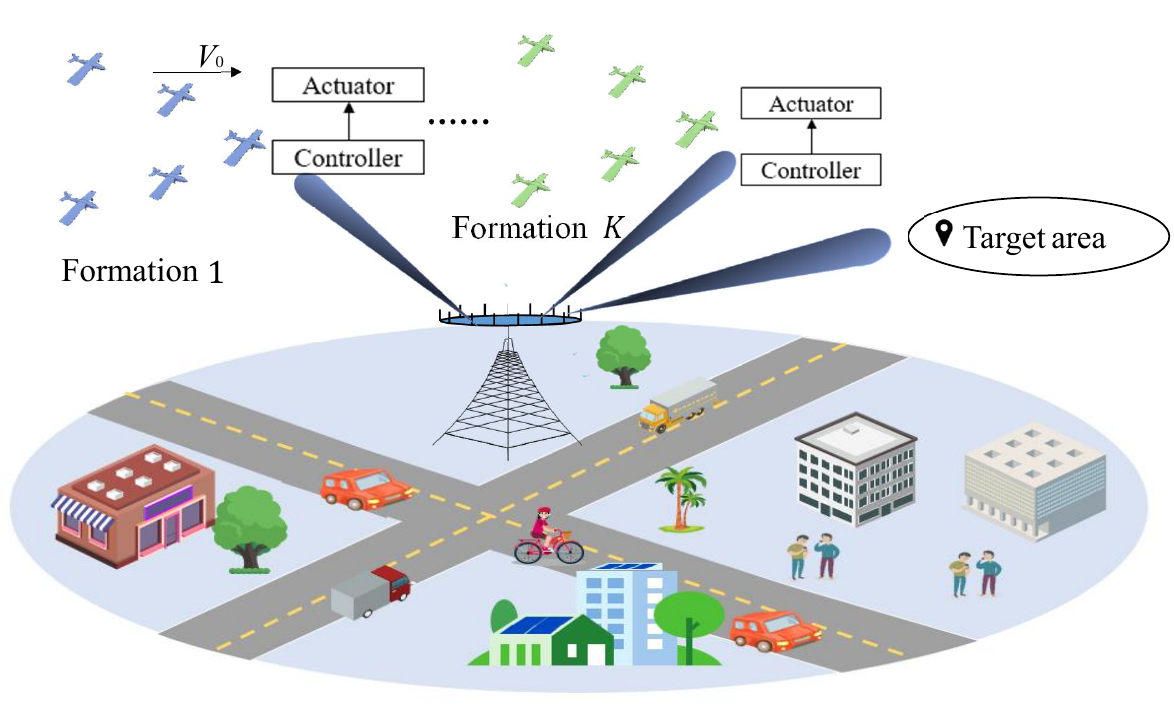}
    \caption{The considered LAWN includes an ISAC-GBS  to continuously monitor the formations and sense the pre-determined target area, providing precise guidance to UAV formations flight.}
    \label{fig1}
 
\end{figure}

As illustrated in Fig. \ref*{fig1}, we consider a typical LAWN involving a dual-functional GBS equipped with \(N_s\) vertically placed antennas relative to the horizontal plane, and \(K\) UAV formations with each consisting of \(M_k\) single-antenna fixed-wing UAVs, \(1 \leq k \leq K\). The two sets are denoted as $\mathcal{K}$ and $\mathcal{M}_k$, respectively. These UAV formations are deployed to navigate a predefined target area for mission-critical tasks, such as search and rescue operations. The ISAC-GBS continuously monitors the formations and senses the target area, providing precise guidance for UAV formation flights. Specifically, the GBS transmits the observed information to the UAV controller, which generates control commands to UAV actuators, e.g., adjusting flight heading, speed, and attitude. Each formation operates independently and is led by a single designated leader. 

To facilitate system design, we focus on a time period of length \( T \), which is uniformly divided into \( N \) discrete time slots, denoted by $\mathcal{N}$. The interval between two consecutive time slots is denoted by \( \Delta t > 0 \), and is chosen to be sufficiently small such that the UAV's position remains constant within each time slot. Without loss of generality, we consider a three-dimensional Cartesian coordinate system, where the \( m \)-th UAV of the \( k \)-th formation (\( m \in \mathcal{M}_k\) and \( k \in \mathcal{K} \)) in the \( n \)-th time slot (\( n \in \mathcal{N}\)) is represented by \( \mathbf{q}_{m,k}[n]=[x_{m,k}[n], y_{m,k}[n], H]^\mathrm{T} \in \mathbb{R}^{3 \times 1} \). Here, the vector \( \mathbf{z}_{m,k}[n]=[x_{m,k}[n], y_{m,k}[n]]^\mathrm{T} \) represents the UAV's horizontal coordinates, while \( H \) denotes the fixed altitude of the formation flight, which can be appropriately selected to accommodate terrestrial infrastructures. We further denote the time-varying position of the \( k \)-th formation leader at the \( n \)-th time slot as $ \mathbf{q}'_{k}[n]=[x'_{k}[n], y'_{k}[n], H]^\mathrm{T} \in \mathbb{R}^{3 \times 1} $.  The location of the ISAC-GBS is fixed at \( \mathbf{b}=[0,0,0]^\mathrm{T} \).
\subsection{Sensing, Communication, and Control Model}
In this work, the UAVs within a formation are assumed to be close sufficiently and exhibit similar behaviors, allowing the GBS to treat the formation as a single entity. This assumption will be further discussed later. The GBS transmits communication signals only to the formation leader, and the remaining followers receive control commands via direct communication with the leader or their neighbors. As stated in \cite{cheng2024networked,9916163}, the GBS sends joint communication and dedicated sensing signals to perform simultaneous downlink communication and target sensing. \revise{Denote $s_k[n]$  and $\mathbf{s}_\mathrm{d} [n]\in \mathbb{C}^{N_s\times1}$ as the transmitted communication signals from the GBS to the leader of $k$-th formation and the dedicated radar sensing signal vector in the $n$-th time slot, respectively.} Note that the signals $s_k[n]$ are mutually independent and follow the circularly symmetric complex Gaussian (CSCG) distribution with zero mean and unit variance, i.e., $s_k[n]\sim \mathcal{CN} (0, 1)$. Similarly, $\mathbf{s}_\mathrm{d}$ is with zero mean and covariance matrix $\mathbf{C}_\mathrm{d}=\mathbb{E}\left\{ \mathbf{s}_\mathrm{d}\mathbf{s}_\mathrm{d}^H \right\}\succeq\boldsymbol{0}$. Then,  the combined signals at the GBS, $\mathbf{s}[n]$, can be written as 
\begin{align}
 \mathbf{s}[n]=\sum_{k=1}^K\mathbf{w}_k[n]s_k[n]+\mathbf{s}_\mathrm{d}[n],\quad\forall n\in\mathcal{N},
\end{align}
where $\mathbf{w}_k[n] \in \mathbb{C}^{N_s\times 1}$ is the transmit beamformer corresponding to the leader of formation $k$ to be designed. Consequently, the covariance matrix of $\mathbf{s}[n]$ can be expressed as 
$    \mathbf{C}[n]=\sum_{k=1}^K\mathbf{w}_k[n]\mathbf{w}^H_k[n]+\mathbf{C}_\mathrm{d}[n].$
As widely assumed in the literature \cite{10168298,8119562,7888557}, the channels from the GBS to UAV formations are dominated by LoS links, and the Doppler effects caused by UAV mobility are well compensated. Then, the channel vector from the GBS to the leader of the formation $k$
is given by 
\begin{align}
    \mathbf{h}_k(\mathbf{q}'_k[n])=\sqrt{\frac{\rho_0}{||\mathbf{q}'_k[n]-\mathbf{b}||^2}}\mathbf{a}(\mathbf{q}'_k[n]), \label{channelgain}
\end{align}
in which $\rho_0$ represents the channel  gain at a unit reference distance and   $\mathbf{a}(\mathbf{q}'_k[n])$ denotes the corresponding steering vector, given by 
\begin{align}
  \mathbf{a}(\mathbf{q}'_k[n])=\left[1, e^{j 2 \pi \frac{d}{\lambda} \cos \theta\left(\mathbf{q}'_k[n] \right)}, \ldots, e^{j 2 \pi \frac{d}{\lambda}(N_s-1) \cos \theta\left(\mathbf{q}'_k[n] \right)}\right]^{\mathrm{T}}, \label{steering}
\end{align}
where $d$ and $\lambda$ denote the spacing between two adjacent antennas and the carrier wavelength, respectively. $ \theta\left(\mathbf{q}'_k[n] \right)$ is the angle of departure (AoD) associated with the leader of the formation $k$, expressed as $\theta\left(\mathbf{q}'_k[n]\right)=\arccos \frac{H}{{\left\|\mathbf{q}'_k[n]-\mathbf{b}\right\|}}.$
 At the UAV side, the received SINR at the leader of the formation $k$ is 
{\small \begin{align}
\mathrm{SINR}_k[n]=\frac{\left|\mathbf{h}^\mathrm{H}_k[n]\mathbf{w}_k[n] \right|^2}{\sum\limits_{{i \in \mathcal{K}\backslash k}}\left|\mathbf{h}_k^\mathrm{H}[n] \mathbf{w}_i[n]\right|^2 +\mathbf{h}_k^\mathrm{H}[n]\mathbf{C}_\mathrm{d}[n]\mathbf{h}_k[n]+\sigma_k^2},
\end{align}}%
where $\mathbf{h}_k[n]$ is the abbreviation of $\mathbf{h}_k(\mathbf{q}'_k[n])$ and $\sigma_k^2$ corresponds to the received noise power. Accordingly, the achievable rate of the leader of the formation $k$ at time slot $n$ is given by $R_k[n]=W\log(1+\mathrm{SINR}_k[n])$, with $W$ being the bandwidth. 

We proceed by investigating the radar sensing within the designated target area. For ease of exposition, the GBS is assumed to sense the target area via a finite number of $J$ sampled points with a fixed height $H$ \cite{cheng2024networked}, denoted by $\mathcal{J}=\{1,2,\cdots,J\}$. The 3D location of the sampled point $j \ (j\in\mathcal{J})$ is expressed as $\mathbf{t}_j \in \mathbb{R}^{3\times 1}$. Relying on (\ref{steering}), the corresponding AoD $\theta(\mathbf{t}_j)$ from the GBS for sensing the sampled point $j$ can be obtained directly. Moreover, we consider that the GBS performs sensing by leveraging both the communication and the radar sensing signals, such that the transmit beampattern gain towards the sampled point $j$ is represented by 
\begin{align}
    \Gamma_j[n]=\mathbf{a}^\mathrm{H}(\mathbf{t}_j) \left(\sum_{k=1}^K\mathbf{ w}_k[n]\mathbf{ w}_k^\mathrm{H}[n]+\mathbf{ C}_\mathrm{d}[n]\right)\mathbf{a}(\mathbf{t}_j). \label{beampattern}
\end{align}
\revise{
The transmit beam pattern gain $\Gamma_j[n]$ serves as a fundamental sensing metric, as it directly determines the received SNR at the sampled points and thus critically affects both detection performance and estimation accuracy \cite{9916163}. Therefore, we adopt the constraint in (\ref{beampattern}) as a practical surrogate to regulate sensing quality.} The GBS sends the observations with respect to the target area to the UAV controller, which then generates optimal control policies.  Without loss of generality, we model the control object as a linear time-invariant system \cite{10050119,8693967}, and the discrete-time system equation of the $k$-th formation\footnote{Here, we assume that the UAVs within one formation have similar behaviors and thus share an identical system equation and control policy.} can be expressed as 
\begin{equation} \mathbf{x}_{k}[n+1]=\mathbf{A}_k\mathbf{x}_{k}[n]+\mathbf{B}_k\mathbf{u}_{k}[n]+\mathbf{v}_{k}[n], \label{sys}
\end{equation}
where $ \mathbf{x}_{k}[n] \in \mathbb{R}^{n_1 \times 1}$ represents the $n_1$-dimensional system state, $\mathbf{u}_{k}[n] \in \mathbb{R}^{n_2 \times 1}$ denotes the $n_2$-dimensional system input, and $\mathbf{v}_{k}[n] $ is the state transition noise with zero mean and covariance matrix $ \boldsymbol{\Sigma}_{\mathrm{v},k}\in \mathbb{R}^{n_1 \times n_1}$. In (\ref{sys}), the matrices $\mathbf{A}_k \in \mathbb{R}^{n_1 \times n_1}$ and $\mathbf{B}_k \in \mathbb{R}^{n_1 \times n_2}$  are deterministic state transition matrix and system input matrix, respectively. Furthermore, the $l$-dimensional observed output can be characterized as 
\begin{align}
    \mathbf{y}_{k}[n]=\mathbf{G}_k\mathbf{x}_{k}[n]+\boldsymbol{\omega}_{k}[n], \label{stateobs}
\end{align}
where $\mathbf{G}_k \in \mathbb{R}^{l\times n_1}$ is a fixed matrix and $\boldsymbol{\omega}_{k}[n]\sim \mathcal{CN}(\mathbf{0},\mathbf{\Sigma}_{\omega,k})$ with $\mathbf{\Sigma}_{\omega,k}\in \mathbb{R}^{l\times l} $ being the covariance matrix. To evaluate the control performance,  the LQR cost function is commonly selected, which, at time slot $n$, is written as 
{\begin{align}
\nonumber \mathrm{LQR}_k[n] \triangleq &\mathbb{E}\bigg[\sum_{i=1}^{n-1}\left(\mathbf{x}_{k}^{\mathrm{T}} [i]\mathbf{Q}_k \mathbf{x}_{k}[i]+\mathbf{u}_{k}^{\mathrm{T}}[i] \mathbf{R}_k \mathbf{u}_{k}[i]\right)\\
 &+\mathbf{x}_{k}^{\mathrm{T}} [n]\mathbf{Q}_{1,k} \mathbf{x}_{k}[n]\bigg], \label{lqr}
\end{align}}
where $\mathbf{Q}_k \succeq \mathbf{0}$ and $\mathbf{Q}_{1,k} \succeq \mathbf{0} $ are state cost square matrices of dimension $n_1$, and $\mathbf{R}_k \succeq \mathbf{0}$
is the control cost square matrix of dimension $n_2$. In (\ref{lqr}), $\mathbf{x}_{k}^{\mathrm{T}} [i]\mathbf{Q}_k \mathbf{x}_{k}[i]$ penalizes deviations of the system state from the desired state, while $\mathbf{u}_{k}^{\mathrm{T}}[i] \mathbf{R}_k \mathbf{u}_{k}[i]$ represents the control power consumption. 
The optimal control policy aims to minimize the LQR cost, maintaining the state close to the desired state $\mathbf{0}$ with less control power.

\subsection{Aerodynamics Model}
\begin{figure}
    \centering
\includegraphics[width=0.8\linewidth]{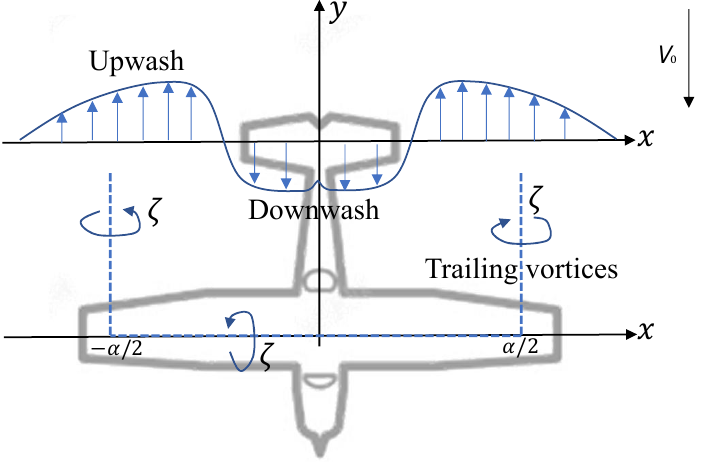}
    \caption{ Horseshoe vortex model}
    \label{voritces}
\end{figure}
As an initial study on UAV formations in the ISAC network, we consider a simplified scenario, where the formations maintain a constant flight direction (e.g., along the negative $y$-axis) and velocity $V_0$. Next, we focus on the force analysis with a single UAV, which can be decomposed into four components, i.e., the lift, gravity, thrust, and drag. Given fixed flying altitude and velocity, the net force should be zero, i.e., the pairs of lift and gravity, thrust, and drag are equally opposite forces. Consequently, the flight power consumption is determined by overcoming the drag, which typically consists of parasite drag and induced drag. The parasite drag is given by $F_\mathrm{p}=\frac{1}{2} cs_0V^2_0$ with $c$ and $s_0$ being the pre-determined system parameters \cite{9322373}, while the induced drag is caused by the wingtip vortex and the downwash effect \cite{weimerskirch2001energy}. Specifically, a UAV flying along the negative $y$-axis generates two semi-infinite trailing vortices with opposite directions at the wingtip due to air interaction. To theoretically characterize vortices, the horseshoe model is extensively employed \cite{higdon1978induced,prandtl1934fundamentals}, as illustrated in Fig. \ref{voritces}, where $\alpha=\beta\pi/4$ denotes the distance between the two vortices with $\beta$ being the UAV's wingspan, and $\zeta$ represents vortex circulation in m$^2$/s. The semi-infinite vortex line then generates induced velocities in its proximal location $[x,y]^T$. According to Biot–Savart’s law \cite{anderson2011ebook,milne1973theoretical}, the velocities $[x,y]^T$ caused by a differentiable vortex line $\mathrm{d}\mathbf{l}$  can be obtained by
\begin{equation}
    v_\mathrm{in}(x,y)=\frac\zeta{4\pi}\oint \frac{ {\mathrm{d}\mathbf{l}}\times {\mathbf{r}}}{| {\mathbf{r}}|^3},
\end{equation}
where $\mathbf{r}$ is the vector from $\mathrm{d}\mathbf{l}$ to $[x,y]^T$. To simplify mathematical derivations, the NASA–Burnham–Hallock model is adopted to approximate $v_\mathrm{in}$, similar to \cite{Binetti2003,5699944}. Moreover, the induced velocity is assumed to suffer from a Gaussian decay along the $y$-axis \cite{5413240}, thus the above equation becomes 
\begin{equation}
\begin{aligned}&v_{\mathrm{in}}(x,y)=\frac{\zeta}{2\pi}\frac{x}{r_c^2+x^2}\left(1+\frac{y}{\sqrt{(\beta/2)^2+y^2}}\right)e^{-\frac{(y-\mu)^2}{2\sigma_0}},\end{aligned} \label{inducedv}
\end{equation}
 where $r_c$ is a known system parameter. $\mu$ and $\sigma_0$ are the mean and standard deviation corresponding to the Gaussian decay, respectively. Consequently, the average induced velocity over the wingspan can be straightforwardly obtained by 
 \begin{equation}
      u_0(x,y)=\frac1 \beta\int_{x-\beta/2}^{x+\beta/2}v(\eta,y) d\eta, 
 \end{equation}
which is further particularized as (\ref{avgup}) (at the top of the next page).
 \begin{figure}[t]
    \centering
\includegraphics[width=0.8\linewidth]{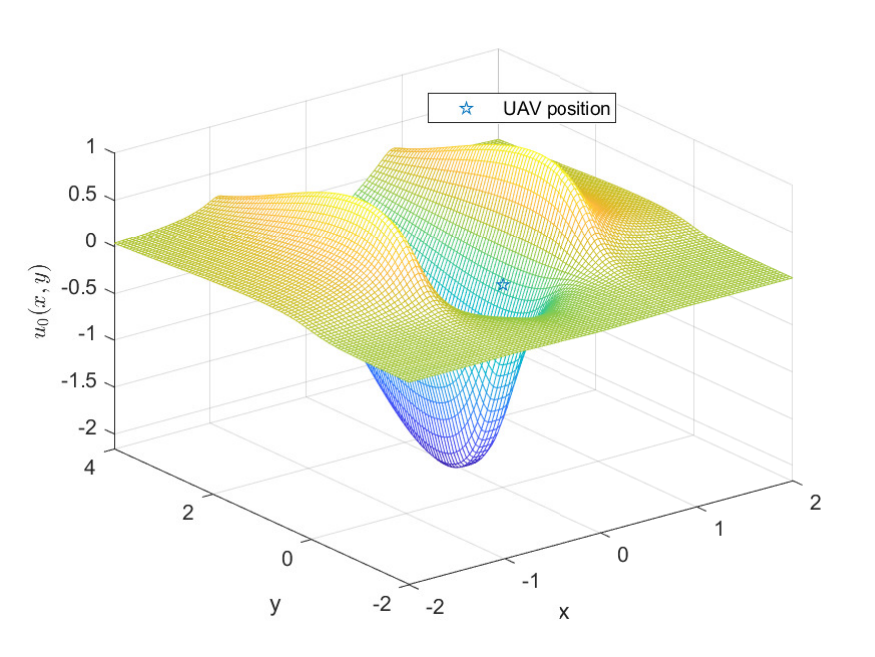}
    \caption{The resultant average induced velocity of a single UAV with horizontal location $[0,0]^T$ m with $\beta=1$ and $r_c=0.1$. }
    \label{upwash}
\end{figure}
 \begin{figure*}[h]
    \begin{align}u_{0}(x,y)=\frac{\zeta}{4\pi \beta}\cdot\left(\ln\frac{(x-\alpha/2+\beta/2)^2+r_c^2}{(x-\alpha/2-\beta/2)^2+r_c^2}-\ln\frac{(x+\alpha/2+\beta/2)^2+r_c^2}{(x+\alpha/2-\beta/2)^2+r_c^2} \right) \left(1+\frac{y}{\sqrt{(\beta/2)^2+y^2}}\right)e^{-\frac{(y-\mu)^2}{2\sigma_0}}    \label{avgup}
   \end{align}    \hrule
\end{figure*}
In general, (\ref{avgup}) consists of upward velocity (upwash) which is generated near the wingtips, and downward velocity (downwash) which is formed near the center of the UAV. To illustrate this, we plot in Fig. \ref{upwash} the generated average induced velocity of a single UAV positioned at a horizontal location $[0,0]^T$ m, flying along $(-y)$-axis direction with a wingspan of $\beta=1$ m. As observed, the upwash zone mainly occurs near the wingtip, while the downwash zone is positioned under the main body. When a trailing UAV flies within the upwash zone created by another UAV's wingtips, it can exploit this upward airflow to effectively reduce the lift required, reducing both the induced drag and energy consumption. By contrast, entering the downwash zone would increase the induced drag, resulting in greater power expenditure. Denote $\Bar{u}$ and $F_l$ as the average upwash at each wing and the lift, respectively, we can obtain a drag reduction at each wing given by $\Delta F_\mathrm{d}=F_l\Bar{u}/V_0$ and a corresponding power reduction $F_l\Bar{u}$ \cite{HUMMEL1983321}. Consequently, the aerodynamic principle of energy-saving formation flight lies in the fact that each UAV flies in the maximum upwash zone generated collectively by the UAVs. For example, the maximum upwash in Fig. \ref{upwash} is located at $[\pm 0.9091, 1.0303 ]^T$ m, which indicates that a trailing UAV should maintain a separation of approximately one wingspan from the lead UAV in both $x$-axis and $y$-axis. To this end, we focus on the formation process by exploiting the upwash rather than directly analyzing the power consumption.

  \vspace{-0.3cm}
\section{ATC Diffusion LMS-Based Energy-Saving Formation} \label{sec3}
In this section, we address the energy-saving formation problem leveraging the ATC diffusion LMS algorithm while considering the upwash effects. Given the assumption of independent formation operations, we concentrate on a single UAV formation, while ensuring the approach can be similarly applied to other formations. Based on (\ref{avgup}), we proceed to analyze the upwash generated by the whole formation. Consider the $m$-th UAV of formation $k$ with horizontal location $\mathbf{z}_{m,k}[n]=\left[x_{m,k}[n], y_{m,k}[n] \right]^T$, the observed total upwash is contributed by the remaining UAVs, which is
\begin{align}
 \nonumber  & u_\mathrm{tol}(x_{m,k}[n], y_{m,k}[n])=\\ &\sum_{m_1\in \mathcal{M}_k \backslash m} u_0(x_{m,k}[n]-x_{m_1,k}[n], y_{m,k}[n]-y_{m_1,k}[n]), \label{upwash_tol}
\end{align}
where we adopt the relative coordinates since the upwash (\ref{avgup}) of each UAV is obtained in their respective coordinate systems. It is noted that the total upwash is time-varying for each UAV and is different across UAVs. 
To establish the formation, it is required to determine a leader. In practice, the leader is typically chosen based on specific operational criteria, such as having the most remaining battery capacity, optimal positional advantage, and superior communication quality with followers. In this work, we simply select the UAV with the smallest value of $y$ as the leader since the formations fly along $-y$ direction. Specifically, the leader index of the $k$-th formation at time slot $n$ is expressed as
\begin{align}
    m_{0}=\mathrm{arg \ \min}_m \ y_{m,k}[n]. \label{leaderder}
\end{align}

As illustrated in Fig. \ref{upwash}, the upwash almost goes to zero far away from the UAV. This implies that the upwash for UAV $m$ is dominated by the nearest preceding UAV, referred to as its reference UAV. In particular, we denote $\mathbf{z}_{m,k}^{\mathrm{ref}}[n]=\left[x_{m,k}^{\mathrm{ref}}[n],y_{m,k}^{\mathrm{ref}}[n]\right]^\mathrm{T}$ as the horizontal location of  reference UAV of UAV $m$, which is determined by 
\begin{equation} 
\begin{aligned}\mathbf{z}_{m,k}^{\mathrm{ref}}[n]&= \underset {\mathbf{z}_{m_1,k}[n]}{\arg\min}\ d_{\kappa}^{2}(\mathbf{z}_{m,k}[n],\mathbf{z}_{m_1,k}[n])\\&\text{s.t.} \ y_{m,k}[n]-y_{m_1,k}[n]>0, m_1\neq m,
\end{aligned} \label{wdis}
\end{equation}
where $d_{\kappa}^{2}(\mathbf{z}_{m,k}[n],\mathbf{z}_{m_1,k}[n])=(x_{m,k}[n]-x_{m_1,k}[n])^2+\kappa(y_{m,k}[n]-y_{m_1,k}[n])^2 $ denotes the weighted distance and $0<\kappa<1$ is the weighting factor. 

\textbf{Remark:} The function of $\kappa$ is to prioritize UAV \( m \) in following a UAV in the \( (-y) \)-direction rather than the \(x\)-direction. For instance, consider a UAV positioned at \([0,0]^T\) with two reference UAV options located at \([0,-1]^T\) and \([1/2,-\sqrt{3}/2]^T\), respectively. Although both UAVs satisfy the Euclidean distance criteria, the UAV located at \([0,-1]^T\) is the preferred based on (\ref{wdis}) since \(1/4 + \kappa 3/4 > \kappa\). Moreover, it is noted that the formation leader does not have a reference UAV given the constraint in (\ref{wdis}).

With the reference UAV at hand, \eqref{upwash_tol} under the reference UAV's coordinate system can be re-expressed as
\begin{align}
  \nonumber & u_\mathrm{tol}(\Delta x_{m,k}[n], \Delta y_{m,k}[n])=\\ &\sum_{m_1\in \mathcal{M}_k \backslash m}u_0(\Delta x_{m,k}[n]-\Delta x_{m_1,k}[n], \Delta y_{m,k}[n]- \Delta y_{m_1,k}[n]), \label{upwash_tol_REF}
\end{align}
where $\Delta x_{m,k}[n]=x_{m,k}[n]-x^{\rm{ref}}_{m,k}[n]$, and $\Delta y_{m,k}[n]=y_{m,k}[n]-y^{\rm{ref}}_{m,k}[n]$.
Fig. \ref{upwash} also shows a pair of maximum upwash points, which means the trailing UAV can either position itself on the left-hand side (LHS) behind its reference UAV or the right-hand side (RHS). For clarity, we denote a binary variable $\Lambda_{m,k}=\{1,-1\}$, where $\Lambda_{m,k}=-1$ represents that the trailing UAV prefers to follow on the RHS of the reference UAV; otherwise, it will locate in the LHS behind its reference UAV, i.e., $\Lambda_{m,k}=1$. Consequently, the optimal relative coordinates of UAV $m$ can be demonstrated by $\left[\Lambda_{m,k}\Delta x^*_{m,k}[n], \Delta y^*_{m,k}[n]\right]^\mathrm{T}$, in which $\Delta x^*_{m,k}$ is assumed to be positive. Intuitively, $\mathbf{\Delta z}^*_{m,k}[n]= \left[\Delta x^*_{m,k}[n], \Delta y^*_{m,k}[n]\right]^\mathrm{T}$ can be obtained by maximizing its observed total upwash \eqref{upwash_tol_REF}, by
\begin{equation}
  \mathbf{\Delta z}^*_{m,k}[n]= \underset{
         \tiny \Delta x_{m,k}[n], 
        \tiny\Delta y_{m,k}[n]
   }{\arg \max} u_\mathrm{tol}(\Delta x_{m,k}[n], \Delta y_{m,k}[n]).\label{upwash_max}
\end{equation}

However, as \eqref{upwash_max} is neither convex nor quasi-convex, it is challenging to achieve a global optimal solution.
Furthermore, the optimal location that maximizes the upwash for each UAV varies across time slots due to the time-dependent nature of the total upwash function. This indicates that even if UAV \(m\) determines its optimal position \(\mathbf{\Delta z}^*_{m,k}[n]\) based on \eqref{upwash_max} and moves from its current location to \(\mathbf{\Delta z}^*_{m,k}[n]\) in the subsequent time slot, this position may not remain optimal for time slot $(n+1)$. Consequently, UAV \(m\) cannot precisely achieve the maximum upwash point; instead, it is constantly moving toward the time-varying optimal location, leading to the failure of UAV formation. \revise{To address this challenge, a centralized node is typically required, which collects upwash observations from all agents and computes the globally optimal formation configuration at each time slot.  While this approach may yield optimal aerodynamic performance, it requires full network knowledge, incurring high computational and communication overhead at the central node and critical reliability risks, as the failure of the central node would compromise the entire formation.}
Given that, we explore a distributed network, where each UAV is only allowed to communicate with its neighbors for exchanging location estimates (rather than upwash observations). The neighbor sets of UAV $m$ within formation $k$ in time slot $n$, denoted by $\mathcal{N}_{m,k}^\mathrm{e}[n]$, include the nearest two UAVs under Euclidean distance criteria as well as itself.

Next, we estimate the optimal locations by utilizing the ATC diffusion LMS approach \cite{5280228,7035117}, which enables decentralized computation through local interactions with neighbors. To start with, we assume that UAV $m$ has a rough estimate of \(\mathbf{\Delta z}^*_{m,k}[n]\), denoted by \(\mathbf{\Delta \hat{z}}_{m,k}[n]\)=\(\mathbf{\Delta z}^*_{m,k}[n-1]\).  To facilitate the UAV formation design, we denote a new upwash observation model expressed as
 \begin{align}
    d_{m,k}[n]= u^\mathrm{max}_{m,k}[n]-u^\mathrm{tol}_{m,k}[n]+\mathbf{f}^{\rm{T}}_{m,k}[n] \mathbf{\Delta \hat{z}}_{m,k}[n]+v_{m,k}[n], \label{upwash_obs}
 \end{align}
where 
$u^\mathrm{max}_{m,k}[n]$ represents the experienced maximum upwash ever before for UAV $m$ and  $u^\mathrm{tol}_{m,k}[n]$ is the abbreviation of $ u_\mathrm{tol}(\Delta x_{m,k}[n], \Delta y_{m,k}[n])$ for notational simplicity. In \eqref{upwash_obs}, $v_{m,k}[n]$ is the corresponding uncorrelated AWGN with zero-mean of variance $\sigma^2_{m,k}$. $\mathbf{f}_{m,k}[n]$ denotes the regression vector, which can be particularized as the gradient of \eqref{upwash_tol_REF} with respect to the  $\left[\Delta x_{m,k}[n], \Delta y_{m,k}[n]\right]^\mathrm{T}$, given by
\begin{align}
\mathbf{f}_{m,k}[n] =\left[\Lambda_{m,k}\frac{\partial u^\mathrm{tol}_{m,k}[n]}{\partial \Delta x_{m,k}[n]}  \ \frac{\partial u^\mathrm{tol}_{m,k}[n]}{\partial \Delta y_{m,k}[n]} \right]^\mathrm{T}.\label{grad}
\end{align}

Our goal is to find $\mathbf{\Delta \hat{z}}_{m,k}[n]$ that can minimize the sum of the mean-square errors across all UAVs, which is
\begin{equation}
\mathbf{\Delta {z}}^*_{m,k}[n]=\arg \min _{\mathbf{\Delta \hat{z}}_{m,k}[n]} \sum_{m=1}^{M_k} \mathbb{E}\left|d_{m,k}[n]-\mathbf{f}^\mathrm{T}_{m,k}[n]\mathbf{\Delta \hat{z}}_{m,k}[n]\right|^2. \label{distributed}
\end{equation}

The above problem can be effectively solved using the ATC diffusion LMS algorithm, which enables each UAV to estimate \(\mathbf{\Delta {z}}^*_{m,k}[n]\) by utilizing the local linearized observation model in \eqref{upwash_obs} and information exchange with its neighboring UAVs. In time slot $n$, all UAVs within the formation, excluding the leader, begin with updating their local estimates \(\boldsymbol{\psi}_{m,k}[n+1]\) using the LMS algorithm, which is expressed as
\begin{align}
   \nonumber \boldsymbol{\psi}_{m,k}[n+1]&\\ =\mathbf{\Delta \hat{z}}_{m,k}[n]&+\mu'_{m,k} \mathbf{f}_{m,k}[n]\left(d_{m,k}[n]-\mathbf{f}^\mathrm{T}_{m,k}[n]\mathbf{\Delta \hat{z}}_{m,k}[n]\right),\label{adaptnotleaser}
\end{align}
where $\mu'_{m,k}$ is the corresponding step-size. Meanwhile, the local estimate for the formation leader is given by
\begin{align}
   \boldsymbol{\psi}_{m_0,k}[n+1]=\mathbf{\Delta \hat{z}}_{m_0,k}[n].  \label{adaptleader}
\end{align}

Subsequently, the UAVs communicate with their neighbors to exchange local estimates, based on which each UAV determines its destinations in the next time slot, by
\begin{align}
    \mathbf{\Delta {z}}^*_{m,k}[n+1]=\sum_{\iota\in \mathcal{N}^e_{m,k}[n]}\omega_{\iota,m,k}\boldsymbol{\psi}_{m,k}[n+1], \label{destination}
\end{align}
with $0\le\omega_{\iota,m,k}\le1$ being the weighting factor, which satisfies that $\sum_{\iota\in \mathcal{N}^e_{m,k}[n]}\omega_{\iota,m,k}=1$ and $\omega_{\iota,m,k}=0$, if ${\iota\notin \mathcal{N}^e_{m,k}[n]}$. 

Apart from the location estimation model, a proper UAV motion model is also essential for a successful formation. Combining \eqref{destination} and the reference UAV location, it is readily to obtain the optimal horizontal global coordinates of UAV $m$ in time slot $n+1$, i.e., $\mathbf{z}^*_{m,k}[n+1]$. Ideally, once the $m$-th UAV obtains the variable $ \mathbf{{z}}^*_{m,k}[n+1]$, it moves to the new coordinates immediately based on $x_{m,k}[n+1]=x_{m,k}^\mathrm{ref}[n]+\Lambda_{m,k}\Delta x^*_{m,k}[n]$ and $y_{m,k}[n+1]=y_{m,k}^\mathrm{ref}[n]+\Delta y^*_{m,k}[n]$. However, due to the limited velocity in practice, the UAVs may not be able to arrive at the desired location in each time slot. To tackle this issue, we introduce a variable $\vartheta \in (0,1)$ to jointly account for current coordinates and the desired coordinates in the next time slot, thus, the UAV motion model can be expressed as (\ref{motion}) shown in the top of this page. In \eqref{motion}, $v^x_{m,k}$ and  $v^y_{m,k}$ correspond to the state transition noises, which obey the Gaussian distributions $\mathcal{CN}(0, \sigma_x^2)$ and  $\mathcal{CN}(0, \sigma_y^2)$, respectively.   
\begin{figure*}[t] 

  \begin{equation} \label{motion}
\left\{\begin{array}{l}
x_{m,k} [n+1]=\vartheta x_{m,k}[n]+(1-\vartheta)\left(x_{m,k}^{\mathrm{ref}}[n]+\Lambda_{m,k}\Delta x^*_{m,k}[n]\right)+v^x_{m,k}, \\
y_{m,k}[n+1]=\vartheta y_{m,k}[n]+(1-\vartheta)\left(y_{m,k}^{\mathrm{ref}}[n]+\Delta y^*_{m,k}[n]\right)-V_0\Delta t+v^y_{m,k}.
\end{array}\right.
\end{equation} 
\hrule
\vspace{-0.5cm}
\end{figure*}
However, such a motion model cannot be directly applied to the formation leader since it does not have a reference UAV. Consequently, we define the motion model for the leader as 
\begin{equation} \label{motionleader}
\left\{\begin{array}{l}
x_{m_0,k}[n+1]=x_{m_0, k}[n]+v^x_{m_0, k}, \\
y_{m_0, k}[n+1]=y_{m_0, k}[n]-V_0\Delta t+v^y_{m_0, k}.
\end{array}\right.
\end{equation}

To sum up, the detailed procedure of the proposed formation approach is outlined in Algorithm \ref{alg1}. Specifically, at the beginning of each time slot, the leader UAV is identified. For UAVs that are not designated as the leader, a reference UAV must first be determined by solving problem \eqref{wdis}, followed by adapting the local estimate using the LMS algorithm via \eqref{adaptnotleaser} based on the observed upwash in (\ref{upwash_obs}). Then, the ultimate location estimate is obtained by incorporating the local estimations from neighbors. By contrast, the leader UAV does not need to select a reference UAV or observe the upwash effects generated by other UAVs, as it occupies the foremost position in the formation and therefore lacks an available upwash zone. Furthermore, the leader UAV needs to execute the `adapt' and `combine' steps to determine the desired location for the next time slot. Afterward, all UAVs adjust their positions according to \eqref{motion} and \eqref{motionleader}, respectively. It is worth noting that Algorithm \ref{alg1} can be applied in parallel across all UAV formations. In fact, these formations naturally self-organize into a `V' shape to maximize the exploitation of upwash, consistent with the phenomena observed in bird flock migration in nature, as further discussed in Sec. \ref{simulation}.
\vspace{-0.5cm}
\section{Control-Oriented Dual-Functional Beamforming Design} \label{sec4}
During the flight, the GBS monitors the UAV formations and senses the target area, followed by sending the observations to the UAV formation leader's controller to generate appropriate control commands. Our goal is to minimize the control LQR cost, which primarily depends on the system state $\mathbf{x}_k[n]$ and the control input $\mathbf{u}_k[n]$ defined in \eqref{lqr}. Since extensive studies have been devoted to the design of detailed control policies $\mathbf{u}_k[n]$, e.g., \cite{10731708,meng2024communication},  this work shifts the focus to explore the impact of channel capacity on the LQR cost in such a wireless control scenario. To this end, we formulate the LQR minimization problem and subsequently develop an efficient algorithm to design the dual-functional beamformer.
\renewcommand{\algorithmicrequire}{\textbf{Initializationn}} 
\renewcommand{\algorithmicensure}{\textbf{Output:}} 
\renewcommand{\thefootnote}{1}

\begin{algorithm}[t]
  \caption{ATC diffusion LMS-based energy-saving UAV formation algorithm} %
  \label{alg1}
  \begin{algorithmic}[1]
    \State
      \textbf{Initialization:} Randomly set the UAV horizontal location $\mathbf{z}_{m,k}[1]$ and $\Lambda_{m,k}$ for each UAV within the formation. Set the total time slots $N$ and the estimate $\mathbf{\Delta z}^*_{m,k}[1]$.  \Repeat
      \State Determine the formation leader UAV based on \eqref{leaderder}. 
    \If{UAV $m$ is not the formation leader}
      \State Determine the reference UAV of UAV $m$ via \eqref{wdis}.
      \State Obtain the upwash observation via (\ref{upwash_obs}) and \eqref{grad}.
    \State \textbf{Adapt step:} Update the local estimate via \eqref{adaptnotleaser}.
    \State \textbf{Combine step:} Communicate with neighbors $\mathcal{N}^\mathrm{e}_{m,k}$ 
    \Statex \hspace{1cm}  and calculate the location $\mathbf{\Delta z}^*_{m,k}[n+1]$ via \eqref{destination}.
    \State Move to the new position based on (\ref{motion}).
    \Else
    \State \textbf{Adapt step:} UAV $m$ is the leader and updates the  \Statex \hspace{1cm} local estimate via (\ref{adaptleader}).
      \State \textbf{Combine step:} Communicate with its neighbors  \Statex \hspace{1cm} $\mathcal{N}^\mathrm{e}_{m,k}$ and calculate $\mathbf{\Delta z}^*_{m,k}[n+1]$ via \eqref{destination}.
          \State Move to the new position based on (\ref{motionleader}).
    \EndIf   
     \State $n\leftarrow n+1$
   \Until($n>N$)
  \end{algorithmic}
\end{algorithm} 
\vspace{-0.65cm}
\renewcommand{\thefootnote}{2}
\subsection{Problem Formulation}
 For the $k$-th UAV formation leader with the system equation in (\ref{sys}) and observation model in \eqref{stateobs}, the average LQR cost across all time slots can be expressed as 
$\bar{l}_k\triangleq \frac{1}{N}\mathbb{E}\bigg[\sum_{i=1}^{N-1}\left(\mathbf{x}_{k}^{\mathrm{T}} [i]\mathbf{Q}_k \mathbf{x}_{k}[i]+\mathbf{u}_{k}^{\mathrm{T}}[i] \mathbf{R}_k \mathbf{u}_{k}[i]\right)+\mathbf{x}_{k}^{\mathrm{T}} [N]\mathbf{Q}_{1,k} \mathbf{x}_{k}[N]\bigg].$ 
 According to \cite{8693967}, the required minimum average communication rate for achieving $\bar{l}_k$ is given by \footnote{It is noted that \eqref{eq:rate_cost_tradeoff-1} is theoretically derived under the assumption of $N\rightarrow \infty$, which is approximated by considering $N$ sufficiently large in this paper.}
 \begin{equation}
\bar{R}_{k,\min}\triangleq h_{k}+\frac{n_1}{2}\log\left(1+\frac{n_1\left(\det(\mathbf{N}_{k}\mathbf{M}_{k})\right)^{\frac{1}{n_1}}}{\bar{l}_{k}-\bar{l}_{k,\min}}\right),\forall k\in\mathcal{K},\label{eq:rate_cost_tradeoff-1}
\end{equation}
where $h_{k}\triangleq\log\mid\det(\mathbf{A}_{k})\mid$
represents the intrinsic entropy rate related to the control stability of the $k$-th leader, while ${\mathbf{N}_{k}}$ and ${\mathbf{M}_{k}}$ can be derived respectively, which are
\begin{align}
\mathbf{N}_{k}&=\mathbf{A}_{k}\boldsymbol{\Sigma}_{k}\mathbf{A}_{k}^\mathrm{T}-\boldsymbol{\Sigma}_{k}+\boldsymbol{\Sigma}_{\mathrm{v},k},\\
\text{and}\ \mathbf{M}_{k}&=\mathbf{S}_{k}\mathbf{B}_{k}\left(\mathbf{R}_{k}+\mathbf{B}_{k}^\mathrm{T}\mathbf{S}_{k}\mathbf{B}_{k}\right)^{-1}\mathbf{B}_{k}^{T}\mathbf{S}_{k},\label{mk}
\end{align}
where $\mathbf{S}_k$ is the solution to the discrete algebraic Riccati equation (DARE), given by 
\begin{equation}
\mathbf{S}_{k}=\mathbf{Q}_{k}+\mathbf{A}_{k}^\mathrm{T}\left(\mathbf{S}_{k}-\mathbf{M}_{k}\right)\mathbf{A}_{k},
\end{equation}
and $\boldsymbol{\Sigma}_{k}$ is the covariance matrix of the estimated system state, expressed as
\begin{equation}
\boldsymbol{\Sigma}_{k}=\mathbf{P}_{k}-\mathbf{K}_{k}\left(\mathbf{G}_{k}\mathbf{P}_{k}\mathbf{G}_{k}^\mathrm{T}+\boldsymbol{\Sigma}_{\omega,k}\right)\mathbf{K}_{k}^\mathrm{T},
\end{equation}
where $\mathbf{P}_{k}$ is the solution to DARE as well, which is
\begin{equation}
\mathbf{P}_{k}=\mathbf{A}_{k}\mathbf{P}_{k}\mathbf{A}_{k}^\mathrm{T}-\mathbf{A}_{k}\mathbf{K}_{k}\left(\mathbf{G}_{k}\mathbf{P}_{k}\mathbf{G}_{k}^\mathrm{T}+\boldsymbol{\Sigma}_{\omega,k}\right)\mathbf{K}_{k}^\mathrm{T}\mathbf{A}_{k}^\mathrm{T}+\boldsymbol{\Sigma}_{\mathrm{v},k},
\end{equation}
with $\mathbf{K}_{k}$ being the steady state Kalman filter gain, written as
\begin{equation}
\mathbf{K}_{k}=\mathbf{P}_{k}\mathbf{G}_{k}^\mathrm{T}\left(\mathbf{G}_{k}\mathbf{P}_{k}\mathbf{G}_{k}^\mathrm{T}+\boldsymbol{\Sigma}_{\omega,k}\right)^{-1}.
\end{equation}
Furthermore, $\bar{l}_{k,\min}$ in \eqref{eq:rate_cost_tradeoff-1} is the minimum LQR cost that can be achieved in the absence of communication rate constraints, such that we have $\bar{l}_{k,\min}=\textrm{tr}\left(\boldsymbol{\Sigma}_{\mathrm{v,}k}\mathbf{S}_{k}\right)+\textrm{tr}\left(\boldsymbol{\Sigma}_{k}\mathbf{S}_{k}\mathbf{A}_{k}^\mathrm{T}\mathbf{M}_{k}\mathbf{A}_{k}\right)$ \cite{8693967}.
 As observed, \eqref{eq:rate_cost_tradeoff-1} is monotonically decreasing with respect to $\bar{l}_k$, which reveals the trade-offs between communication rate and LQR cost, i.e., generating more precise control commands at the controller needs receiving more observation information from the GBS. Based on \eqref{eq:rate_cost_tradeoff-1}, we then have the following LQR cost minimization problem: 
\begin{subequations}\label{P1}
\begin{align}
\min_{\mathbf{w}_k[n],\mathbf{C}_\mathrm{d}[n]}& \quad  \max_{\bar{l}_{k},k\in\mathcal{K}}\:\bar{l}_{k}\label{P1o}\\
\hspace{-0.5cm} \mathrm{\textrm{s.t.}} \quad   \sum_{n\in \mathcal{N}}&\Gamma_j[n] \ge \Gamma_\mathrm{th}d^2(\mathbf{t}_j,\mathbf{b}), \forall j, \label{P1-C1}\\
  \quad \frac{1}{N}&\sum_{n\in\mathcal{N}} R_k[n]\ge \bar{R}_{k,\min}, \forall k\in\mathcal{K},\label{P1-C2}\\
\sum_{k\in \mathcal{K}}& ||\mathbf{w}_k[n]||^2+\operatorname{tr}(\mathbf{C}_\mathrm{d}[n]) \leq P_{\max}, \forall n\in \mathcal{N},\label{P1-C3}
\end{align}
\end{subequations}
where $\Gamma_\mathrm{th}$ is a predetermined constant associated with the required sensing performance, $d(\mathbf{t}_j,\mathbf{b})=||\mathbf{t}_j-\mathbf{b}||$ is the distance between the GBS and the $j$-th sampled sensing point, and $P_{\max}$ denotes the maximum available power budget. The objective function \eqref{P1o} is introduced to guarantee control fairness, where we minimize the maximum LQR cost across all formations. Constraint \eqref{P1-C1} implies that the transmit beampattern gain at target points cannot fall below a certain threshold, which is proportional to the square of the distance $d(\mathbf{t}_j,\mathbf{b})$. (\ref{P1-C2}) is imposed to account for the communication rate-LQR cost trade-off shown in \eqref{eq:rate_cost_tradeoff-1}. It is evident that problem \eqref{P1} is non-convex due to constraint (\ref{P1-C2}) and thus, is challenging to tackle directly using conventional convex optimization methods.
\subsection{Efficient Solutions}
In this subsection, we propose an efficient algorithm to deal with problem \eqref{P1} by employing SCA techniques for obtaining a sub-optimal solution. \revise{It is noteworthy that $\mathbf{S}_k$ and $\mathbf{P}_k$ are undetermined due to the DAREs, which can be iteratively addressed respectively, by
{\small \begin{align}\label{iter}
\mathbf{S}^{r+1}_{k}&=\mathbf{Q}_{k}+\mathbf{A}_{k}^\mathrm{T}\left(\mathbf{S}^{r}_{k}-\mathbf{M}^r_{k}\right)\mathbf{A}_{k}, \ \text{and}\\ \nonumber
\mathbf{P}^{r+1}_{k}&=\mathbf{A}_{k}\mathbf{P}^{r}_{k}\mathbf{A}_{k}^\mathrm{T}-\mathbf{A}_{k}\mathbf{K}_{k}\left(\mathbf{G}_{k}\mathbf{P}^{r}_{k}\mathbf{G}_{k}^\mathrm{T}+\boldsymbol{\Sigma}_{\omega,k}\right)\mathbf{K}_{k}^\mathrm{T}\mathbf{A}_{k}^\mathrm{T}+\boldsymbol{\Sigma}_{\mathrm{v},k},%
\end{align} }}%
where $r$ is the iteration index. The iterations terminate when$||\mathbf{S}^{r+1}_{k}-\mathbf{S}^{r}_{k} ||\le\varepsilon_0$ and $||\mathbf{P}^{r+1}_{k}-\mathbf{P}^{r}_{k} ||\le\varepsilon_0$, with $\varepsilon_0$ being the convergence tolerance. Subsequently, we observe that the original problem can be equivalently converted into the following problem by introducing an auxiliary variable $\eta$, which is
\begin{subequations}\label{P2}
\begin{align}
\min_{\{\mathbf{w}_k[n],\mathbf{C}_\mathrm{d}[n],\bar{l}_{k},\eta\}} \hspace{-0.2cm}& \eta \\
\hspace{-0.5cm} \mathrm{\textrm{s.t.}} \quad  \bar{l}_{k}&\le\eta,\quad  \forall k\in\mathcal{K},\label{P2-c1}\\ \sum_{n\in \mathcal{N}}&\Gamma_j[n] \ge \Gamma_\mathrm{th}d^2(\mathbf{t}_j,\mathbf{b}), \forall j\in \mathcal{J}, \label{P2-C2}\\
  \quad \frac{1}{N}&\sum_{n\in\mathcal{N}} R_k[n]\ge \bar{R}_{k,\min}, \forall k\in\mathcal{K},\label{P2-C3}\\
\sum_{k\in \mathcal{K}}&||\mathbf{w}_k[n]||^2+\operatorname{tr}(\mathbf{C}_\mathrm{d}[n]) \leq P_{\max}.\label{P2-C4}
\end{align}
\end{subequations}

Although problem \eqref{P2} is more manageable, it remains non-convex. Moreover, we observe that the variable $\bar{l}_k$ appears only in constraints \eqref{P2-c1} and \eqref{P2-C3}. Consequently, a two-step approach can be proposed by initially addressing $\bar{l}_k$ concerning arbitrary beamformers $\mathbf{w}_k[n]$ and $\mathbf{C}_\mathrm{d}[n]$. Then, we are capable of optimizing the variables $\mathbf{w}_k[n]$ and $\mathbf{C}_\mathrm{d}[n]$ by excluding constraints \eqref{P2-c1} and \eqref{P2-C3}. To this end, with fixed $\mathbf{w}_k[n]$ and $\mathbf{C}_\mathrm{d}[n]$,  we have the following sub-problem with respect to $\bar{l}_k$:
\begin{subequations}\label{P3}
\begin{align}
 \min_{\{\bar{l}_{k},\eta\}} & \quad \eta \\
\hspace{-0.5cm} \mathrm{\textrm{s.t.}} \quad  \bar{l}_{k}&\le\eta,\quad  \forall k\in\mathcal{K},\label{P3-c1}\\ 
  \quad \frac{1}{N}&\sum_{n\in\mathcal{N}} R_k[n]\ge \bar{R}_{k,\min}, \forall k\in\mathcal{K}.\label{P3-C3}
\end{align}
\end{subequations}
 Then, we have the following proposition:

 \textbf{Proposition 1:} \textit{Problem \eqref{P3} is convex and the optimal solution, given by 
 \begin{align}
\bar{l}_k^{\star}=\frac{n_1\left(\operatorname{det} (\mathbf{N}_k \mathbf{M}_k)\right)^{\frac{1}{n_1}}}{\exp({{2R'_{k}}/{n_1}})-1}+\bar{l}_{k,\min } \label{lstar}
 \end{align}
 with $R'_k=\ln2 \left(\frac{1}{N}\sum_{n\in\mathcal{N}} R_k[n]-h_k \right)$. The optimal objective value can be further derived by $\eta=\max\{\bar{l}^*_1, \cdots, \bar{l}^*_K\}$.  }
 
\textit{Proof:} Since the right-hand side (RHS) of \eqref{P3-C3} is monotonically decreasing with respect to $\bar{l}_k$, constraint \eqref{P3-C3} can be equivalently transformed into 
\begin{equation}
\bar{l}_{k}\geq\frac{n_1\left(\det(\mathbf{N}_{k}\mathbf{M}_{k})\right)^{\frac{1}{n_1}}}{2^{\frac{2}{n_1}\left(\frac{1}{N}\underset{n\in\mathcal{N}}{\sum}R_{k}[n]-h_{k}\right)}-1}+\bar{l}_{k,\min},\forall k.\label{rct}
\end{equation}
Substituting \eqref{rct} into \eqref{P3}, we then have 
\begin{equation}\label{P4}
\begin{split}
\min_{\{\bar{l}_{k},\eta\}} & \quad \eta \\
\hspace{-0.5cm} \mathrm{\textrm{s.t.}} \ &\frac{n_1\left(\det(\mathbf{N}_{k}\mathbf{M}_{k})\right)^{\frac{1}{n_1}}}{2^{\frac{2}{n_1}\left(\frac{1}{N}\underset{n\in\mathcal{N}}{\sum}R_{k}[n]-h_{k}\right)}-1}+\bar{l}_{k,\min}\le\bar{l}_{k}\le\eta,\   \forall k.
\end{split}
\end{equation}
It is evident that problem \eqref{P4} is a linear programming (LP) problem and the optimal solution is given by \eqref{lstar}, which completes the proof.  

Based on proposition 1, we can express $\bar{l}_k$ as a function of the dual-functional beamformers, so problem \eqref{P2} can be simplified as 
\begin{subequations}\label{P5}
\begin{align}
&\hspace{-0.8cm}\min_{\{\mathbf{w}_k[n],\mathbf{C}_\mathrm{d}[n],\eta\}}   \eta \qquad \qquad \qquad \qquad \\
 \mathrm{\textrm{s.t.}} \  &\frac{n_1\left(\operatorname{det} (\mathbf{N}_k \mathbf{M}_k)\right)^{\frac{1}{n_1}}}{\exp({{2R'_{k}}/{n_1}})-1}+\bar{l}_{k,\min } \le\eta,\quad  \forall k,\label{P5-c1}\\
& \eqref{P2-C2}, \eqref{P2-C4}.
\end{align}
\end{subequations}
However, problem (\ref{P5}) is intractable to obtain a global optimal solution due to the complicated constraint \eqref{P5-c1}. To address this problem, we propose a computationally efficient algorithm to achieve a sub-optimal solution by harnessing the SCA technique in what follows. To facilitate the solution design, let us denote two slack variables $\mathbf{H}_k[n]=\mathbf{h}_k[n]\mathbf{h}^\mathrm{H}_k[n]$ and $\mathbf{W}_k[n]=\mathbf{w}_k[n]\mathbf{w}^\mathrm{H}_k[n], \forall k,n$. It is worth mentioning that $\mathbf{W}_k[n]$ is a semi-definite positive (SDP) matrix with $\mathrm{rank}(\mathbf{W}_k[n])\le1$. Consequently, problem (\ref{P5}) can be re-expressed as 
    \begin{subequations}\label{P6}
\begin{align}
&\hspace{-0.45cm}\min_{\{\mathbf{W}_k[n],\mathbf{C}_\mathrm{d}[n],\eta\}}   \eta \qquad \qquad \qquad \qquad \\
 &\hspace{-0.25cm}\mathrm{\textrm{s.t.}}  \sum_{k\in \mathcal{K}} \operatorname{tr}\left(\mathbf{W}_{k}[n]\right)+\operatorname{tr}(\mathbf{C}_\mathrm{d}[n]) \leq P_{\max}, \forall n,\label{P6-C1}\\  
  &\hspace{-0.25cm}\sum_{n\in \mathcal{N}}\mathbf{a}^H(\mathbf{t}_j) \big(\sum_{k=1}^K\mathbf{ W}_k[n]+\mathbf{ C}_\mathrm{d}[n]\big)\mathbf{a}(\mathbf{t}_j)\ge \Gamma_\mathrm{th}d^2(\mathbf{t}_j,\mathbf{b}), \forall j,\label{P6-C2}\\
  &\hspace{-0.25cm}\frac{n_1}{2}\log \left( 1+ \frac{\left(\operatorname{det} (\mathbf{N}_k \mathbf{M}_k)\right)^{\frac{1}{n_1}}}{\eta-\bar{l}_{k,\min}}\right) +h_{k} \le\frac{W}{N} \sum_{n\in \mathcal{N}}\hat{R}_k[n], \forall k,\label{P6-c3}\\
 &\mathrm{rank}(\mathbf{W}_k[n])\le1, \forall k,n,\label{P6-C4}
\end{align}
\end{subequations}
where $\hat{R}_k[n]$ is given by \eqref{rateW} positioned in the top of this page.
{\begin{figure*}[t]
    \begin{equation} \label{rateW}
       \hat{R}_k[n] =\log \left(1+\frac{\operatorname{tr}\left(\mathbf{H}_k[n]\mathbf{W}_k[n]\right)}{\sum\limits_{i \in\mathcal{K}\backslash k} \operatorname{tr}\left(\mathbf{H}_k[n]\mathbf{W}_i[n]\right)+\operatorname{tr}\left(\mathbf{H}_k[n]\mathbf{C}_\mathrm{d}[n]\right)+\sigma_k^2}\right)
    \end{equation} 
    \hrule
\end{figure*}}
As observed, problem (\ref{P6}) is non-convex, and the non-convexity arises from constraints \eqref{P6-c3} and \eqref{P6-C4}. We then discard the rank-one constraint \eqref{P6-C4} to obtain a semi-definite relax (SDR) form of \eqref{P6}, thus, we now only need to address \eqref{P6-c3}. Furthermore, we note that $\eta\ge\bar{l}_{k,\min}$ in \eqref{P6-c3} always holds, and thereby, it can be verified that the LHS of \eqref{P6-c3} is convex concerning $\eta$. The remaining is to handle the RHS of \eqref{P6-c3}, i.e. $\hat{R}_k[n]$, which can be further rewritten as $ \hat{R}_k[n] =\log \left({\sum\limits_{i\in\mathcal{K}} \operatorname{tr}\left(\mathbf{H}_k[n]\mathbf{W}_i[n]\right)+\operatorname{tr}\left(\mathbf{H}_k[n]\mathbf{C}_\mathrm{d}[n]\right)+\sigma_k^2}\right)-\hat{R}^s_k[n]$ with the term {$ 
 \hat{R}^s_k[n]=\log\big(\sum\limits_{i=\mathcal{K}\backslash k} \operatorname{tr}\left(\mathbf{H}_k[n]\mathbf{W}_i[n]\right)+\operatorname{tr}\left(\mathbf{H}_k[n]\mathbf{C}_\mathrm{d}[n]\right)$}$+\sigma_k^2\big)$. We observe that $\hat{R}^s_k[n]$ has a convex subset given the local points $\{\mathbf{W}_k^{r_1}[n],\mathbf{C}_\mathrm{d}^{r_1}[n] \}$ by employing the SCA technique with $r_1$ being the iteration index,  which is
    \begin{align}
\nonumber \hat{R}^s_k[n]&\le\hat{R}^{s,\mathrm{lo}}_k[n]+\sum_{i=1, i \neq k}^K \operatorname{tr}\left(\mathbf{D}_k^{r_1}[n]\left(\mathbf{W}_i[n]-\mathbf{W}_i^{r_1}[n]\right)\right) \\
& +\operatorname{tr}\left(\mathbf{D}_k^{r_1}[n]\left(\mathbf{C}_\mathrm{d}[n]-\mathbf{C}_\mathrm{d}^{r_1}[n]\right)\right) \triangleq \hat{R}^{s,\mathrm{ub}}_k[n], \label{upr}
\end{align}
where $\hat{R}^{s,\mathrm{lo}}_k[n]$ is obtained by substituting $\{\mathbf{W}_k^{r_1}[n],\mathbf{C}_\mathrm{d}^{r_1}[n] \}$ to replace their counterparts $\{\mathbf{W}_k[n],\mathbf{C}_\mathrm{d}[n] \}$ in $\hat{R}^{s}_k[n]$, and $\mathbf{D}_k[n]$ is the first-order derivative, given by 
\begin{equation}
    \mathbf{D}_k^{r_1}[n]=\frac{\log(e) \mathbf{H}_k[n]}{\sum\limits_{i=1, i \neq k}^K \operatorname{tr}\left(\mathbf{H}_k[n]\mathbf{W}_i^{r_1}[n]\right)+\operatorname{tr}\left(\mathbf{H}_k[n] \mathbf{C}_\mathrm{d}^{r_1}[n]\right)+\sigma^2}.
\end{equation}

Plugging \eqref{upr} into $\hat{R}_k[n]$, it is then lower-bounded by 
\begin{align}
      \nonumber  \hat{R}_k[n]&\ge\log \left({\sum\limits_{i\in\mathcal{K}} \operatorname{tr}\left(\mathbf{H}_k[n]\mathbf{W}_k[n]\right)+\operatorname{tr}\left(\mathbf{H}_k[n]\mathbf{C}_\mathrm{d}[n]\right)+\sigma_k^2}\right)\\
        &-\hat{R}^{s,\mathrm{ub}}_k[n]\triangleq\hat{R}^{\mathrm{lb}}_k[n]. \label{lbr}
\end{align}

Subsequently, by taking \eqref{lbr} to replace the RHS of \eqref{P6-c3}, we have the following approximated problem:
    \begin{subequations}\label{P7}
\begin{align}
&\hspace{-0.15cm}\min_{\{\mathbf{W}_k[n],\mathbf{C}_\mathrm{d}[n],\eta\}}   \eta \qquad \qquad \qquad \qquad \\
 &\mathrm{\textrm{s.t.}} 
  \frac{n_1}{2}\log \left( 1+ \frac{\left(\operatorname{det} (\mathbf{N}_k \mathbf{M}_k)\right)^{\frac{1}{n_1}}}{\eta-\bar{l}_{k,\min}}\right) +h_{k} \le\frac{W}{N} \sum_{n\in \mathcal{N}}\hat{R}^{\mathrm{lb}}_k[n], \forall k,  \label{p7-c1}\\
 & \qquad \eqref{P6-C1} \text{ and } \eqref{P6-C2}.
\end{align}
\end{subequations}

Problem \eqref{P7} is now convex and can be efficiently solved by standard solvers, e.g., CVX \cite{cheng2024networked,grant2014cvx}. It is noted the resultant solution may not satisfy the stringent rank-1 constraint. Consequently, further steps are required to obtain a feasible solution. Toward this end,  one can employ the singular value decomposition (SVD) method \cite{9676676} or the Gaussian randomization approach \cite{10014666} to appropriately extract the rank-1 solution from the SDR solution in general. However, these two methods would cause an increase in the objective function value. To overcome this challenge, we have the following proposition:

\textbf{Proposition 2:} \textit{If the obtained solutions of problem \eqref{P7}, say $\{\mathbf{\tilde{W}}_k[n],\mathbf{\tilde{C}}_\mathrm{d}[n]\}$, do not satisfy $\mathrm{rank}(\mathbf{\tilde{W}}_k[n])\le1$, one can re-construct the equivalent solutions via \eqref{reso} while not changing the objective function value, i.e., 
\begin{equation} \label{reso}
\begin{aligned}
{\mathbf{w}}^*_k[n] & =\frac{{\tilde{\mathbf{W}}_k[n]} \mathbf{h}_k[n]}{\sqrt{\operatorname{tr}(\tilde{\mathbf{W}}_k[n]\mathbf{H}_k[n])}}, \\
{\mathbf{W}}^*_k[n] & =\mathbf{w}^*_k[n]\mathbf{w}_k^{*,\mathrm{H}}[n], \\
{\mathbf{C}}_\mathrm{d}^*[n] & =\sum_{k=1}^K \tilde{\mathbf{W}}_k[n]+\tilde{\mathbf{C}}_\mathrm{d}[n]-\sum_{k=1}^K {\mathbf{W}}^*_k[n] .
\end{aligned}
\end{equation}
}  
\textit{Proof:} See \cite{9916163}.

It is evident that ${\mathbf{W}}^*_k[n]$ is a rank-1 matrix. Thus, problem \eqref{P1} can be efficiently solved in an iterative manner, which can guarantee monotonically non-increasing objective values. Moreover, since the objective value is lower-bounded by $\bar{l}_{k,\min}$, problem \eqref{P7} can converge to a finite value.  To sum up, the overall algorithm for solving problem \eqref{P1} is presented in Algorithm \ref{alg2}.
\revise{
We next quantify the computational complexity of the proposed scheme. For \textbf{Algorithm 1}, reference-UAV selection incurs $\mathcal{O}(M_k^{2})$ operations per slot within a single formation, such that the total cost across $N$ slots is dominated by $\mathcal{O}\!\bigl(N\sum_{k=1}^{K}M_k^{2}\bigr)$. As for \textbf{ Algorithm 2},  the SDP imposes a complexity of $\mathcal{O}\!\bigl(((K+1)N_s^2)^{3.5}\bigr)$ by adopting the interior-point method, while the auxiliary DARE updates contribute $\mathcal{O}(n_1^{3})$ and are negligible for $n_1\ll N_s$. Let $T_1$ denote the required number of iterations, such that the complete polynomial complexity approximately becomes $\mathcal{O}\!\bigl(N\sum_{k=1}^{K}M_k^{2}+\bigl(T_1N((K+1)N_s^2)^{3.5}\bigr)\bigr)$.}

\begin{algorithm}[t]
  \caption{Overall algorithm for solving problem \eqref{P1}} %
  \label{alg2}
  \begin{algorithmic}[1]
    \State
      \textbf{Initialization:} initialize the beamformer $\mathbf{W}_k^0[n]$, and $\mathbf{C}_\mathrm{d}^0[n]$. Set the iteration index $r_1$=0.
      \State Solve matrices $\mathbf{S}_k$ and $\mathbf{P}_k$ based on \eqref{iter}.
      \Repeat
      \State Obtain the SDR solutions $\{\mathbf{\tilde{W}}_k[n],\mathbf{\tilde{C}}_\mathrm{d}[n]\}$ by solving 
      \Statex \hspace{0.41cm} problem \eqref{P7} given local points $\{\mathbf{W}_k^{r_1}[n],\mathbf{C}_\mathrm{d}^{r_1}[n]\}$. 
      \State Re-construct the equivalent solutions  $\{\mathbf{{W}}_k^*[n],\mathbf{{C}}_\mathrm{d}^*[n]\}$ 
      \Statex \hspace{0.41cm} based on \eqref{reso}.
      \State $\{\mathbf{W}_k^{r_1}[n],\mathbf{C}_\mathrm{d}^{r_1}[n]\}\leftarrow \{\mathbf{{W}}_k^*[n],\mathbf{{C}}_\mathrm{d}^*[n]\}$.
     \State $r_1\leftarrow r_1+1$
   \Until(Converged)
  \end{algorithmic}  
\end{algorithm} 
\section{Simulation Results} \label{simulation}
  In this section, we conduct extensive simulations to validate the effectiveness of our proposed design. We consider a square area size of $0.5$ km $\times$ $0.5$ km, where the GBS is fixed at $\mathbf{b}=[0,0,0]^\mathrm{T}$ m.  The antenna spacing of GBS is $d=\lambda/2$ with the number of antennas being $N_s=12$. The number of UAV formations is $K=2$, in which the first formation consists of $M_1=19$ UAVs and the second involves $M_2=9$ UAVs. We set the parameters of each UAV as follows: wingspan $\beta=1$ m, vortex separation $\alpha=\pi/4$, vortex radius $r_c=0.1$ m, $\zeta=2$, $\mu=0.7$, and $\sigma_0=4$, respectively \cite{5413240,10679984}. The velocity of each UAV is set to $V_0=5$ m/s with a fixed altitude $H=30$ m and the duration of each time slot corresponds to $\Delta t=0.05$ s. The variances associated with position updates are $\sigma_x^2=\sigma_y^2=2\times 10^{-4} $. The weighting factors are $\kappa=1/3$, $\vartheta=0.5$, and $\omega_{\iota,m,k}=1/3$ if ${\iota\in \mathcal{N}^e_{m,k}}, \forall m,k$. The step size of LMS algorithm is $\mu'_{m,k}=2\times 10^{-3} $. As for control,  the system state dimensions and observation dimensions are both set to $n_1=l=50$. The covariance matrices of the system transition model and observation model are  $\mathbf{\Sigma}_v=0.01\times\mathbf{I}_{n_1}$ and $\mathbf{\Sigma}_\omega=0.001\times \mathbf{I}_l$, respectively. For ease of exposition, we set the input matrix $\mathbf{B}_k$, the observation matrix $\mathbf{G}_k$, and the state cost square matrix $\mathbf{Q}_k$ to identity matrices \cite{10050119}. Moreover, the control cost square matrix $\mathbf{R}_k$ is a zero matrix. The number of sample locations is $J=20$, which are randomly located within a rectangular area with $ x$-coordinates ranging from $15$ m to $85$ m and $ y$-coordinates ranging from $-140$ m to $-130$ m. Unless otherwise specified, the maximum power budget is limited to $P_{\max}=1$ W ($30$ dBm), and the sensing threshold is set to  $\Gamma=0$ dBm. The noise power for each formation leader is set to $\sigma^2_k=-90$ dBm, and the channel power gain at a reference unit distance is $\rho_0=-60$ dB \cite{9916163}. To effectively exploit the upwash, the UAV locations of each formation should be carefully initialized. On the one hand, if the UAVs are far from each other, they may not be capable of feeling the upwash and moving to the desired destinations. On the other hand, if the UAVs are too close to each other, they may suffer from collision risks. To overcome this concern, we randomly initialize the $ x$-coordinates of UAVs ranging from $-M_k\beta/4$ to $M_k\beta/4$ and the $ y$-coordinates ranging from $0$ to $M_k\beta/2$. \revise{Furthermore, we assume that there exists an equal number of UAVs on the LHS and RHS with respect to the leader, which is achieved by randomly setting $\Lambda_{m,k}$ to be $1$ or $-1$ with equal probability. Note that the proposed algorithm is still effective in the situation where the number of LHS and RHS UAVs is not identical.}


\begin{figure*}[t]
  \centering
 \begin{subfigure}{0.67\columnwidth}
  \centering
\includegraphics[width=\linewidth]{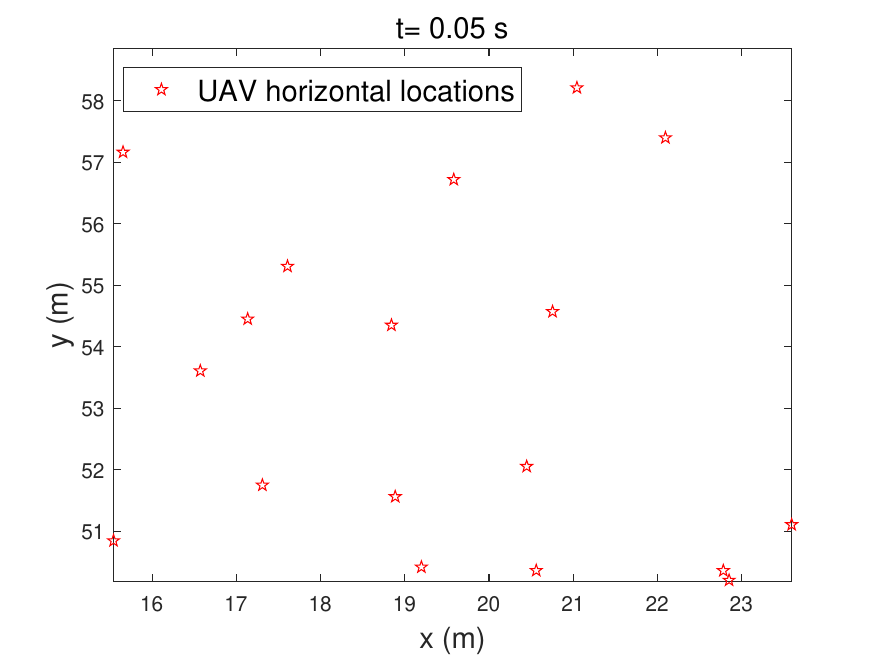}
       \caption{Locations, $t=0.05$ s.}
    \label{figsubber}
  \end{subfigure}%
\begin{subfigure}{0.67\columnwidth}
  \centering
\includegraphics[width=\linewidth]{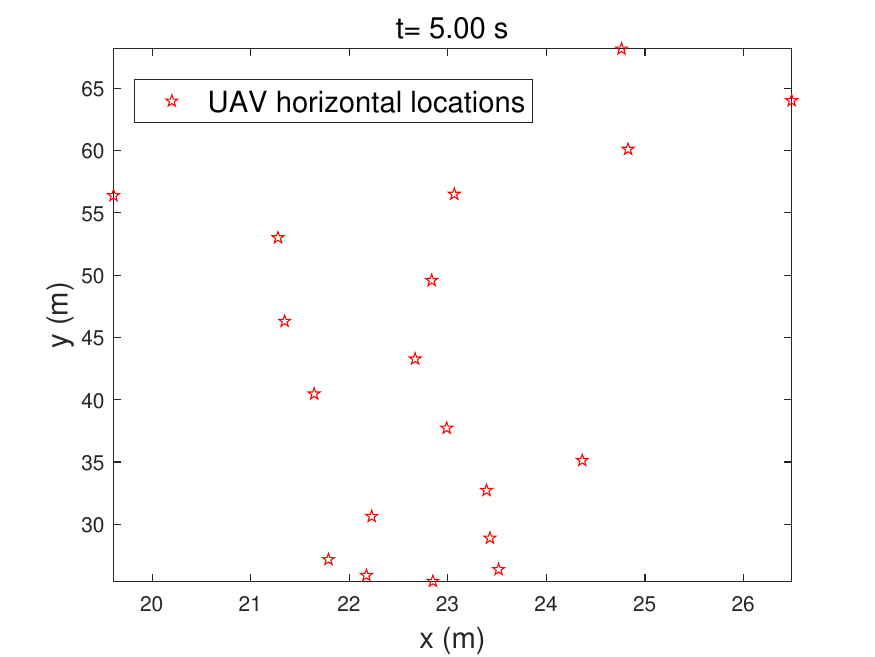}
    \caption{Locations, $t=5$ s.}
    \label{figsubcrb}
  \end{subfigure}
\begin{subfigure}{0.67\columnwidth}
    \centering
\includegraphics[width=\linewidth]{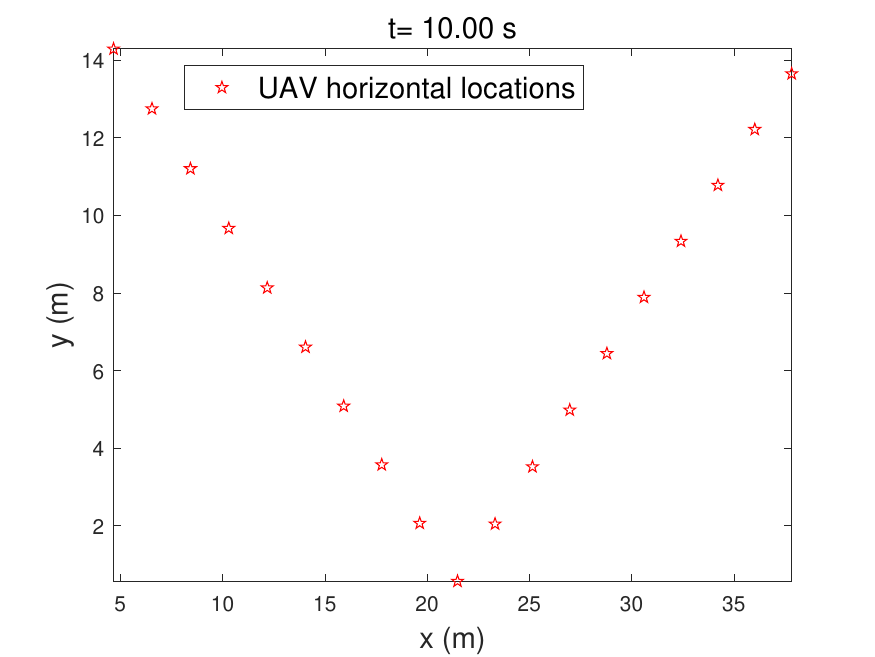}
    \caption{ Locations, $t=10$ s. }
    \label{figsubrate}
  \end{subfigure}
  \begin{subfigure}{0.67\columnwidth}
  \centering
\includegraphics[width=\linewidth]{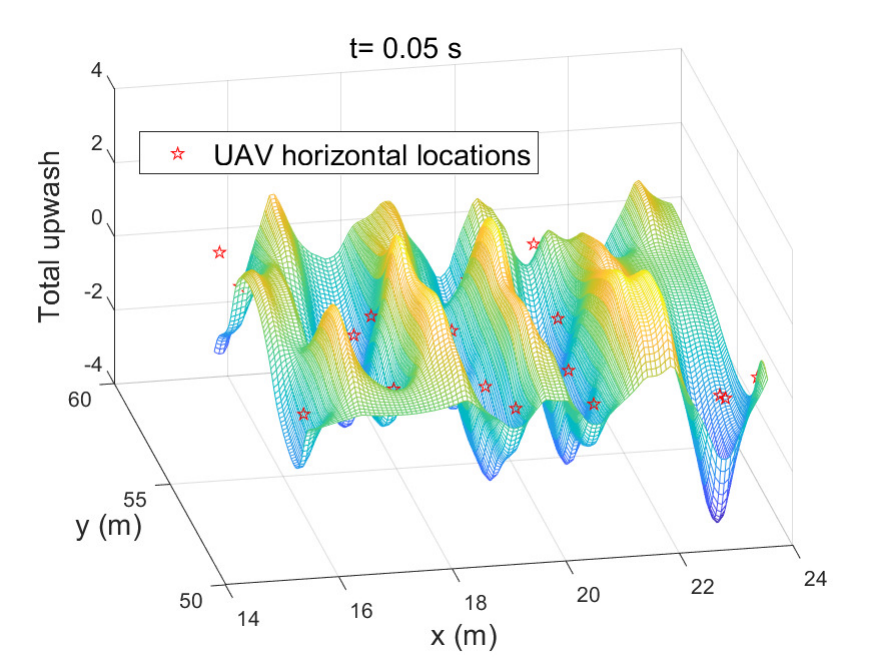}
       \caption{Total upwash, $t=0.05$ s.}
    \label{figsubber}
  \end{subfigure}%
\begin{subfigure}{0.67\columnwidth}
  \centering
\includegraphics[width=\linewidth]{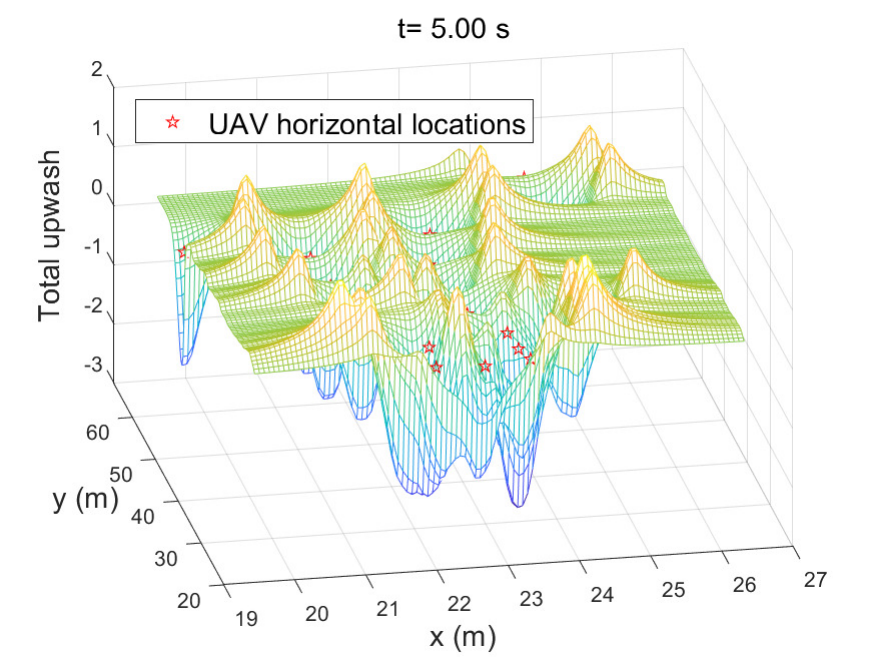}
    \caption{Total upwash, $t=5$ s.}
    \label{figsubcrb}
  \end{subfigure}
\begin{subfigure}{0.67\columnwidth}
    \centering
\includegraphics[width=\linewidth]{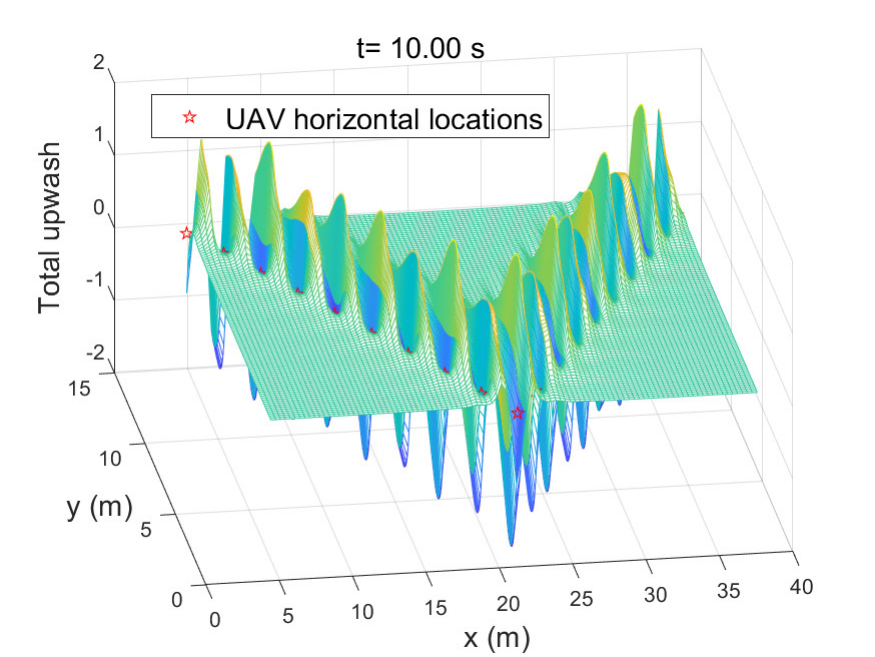}
    \caption{Total upwash, $t=10$ s. }
    \label{upwash19}
  \end{subfigure}
\captionsetup{font=small,name=Fig,labelsep=period,justification=justified,singlelinecheck=false}
  \caption{The horizontal locations and the total upwash of formation 1 including $19$ UAVs in various time slots. }
   \label{figsys}
  \end{figure*}

\begin{figure*}[t] 
  \centering
 \begin{subfigure}{0.5\columnwidth}
  \centering
\includegraphics[width=\linewidth]{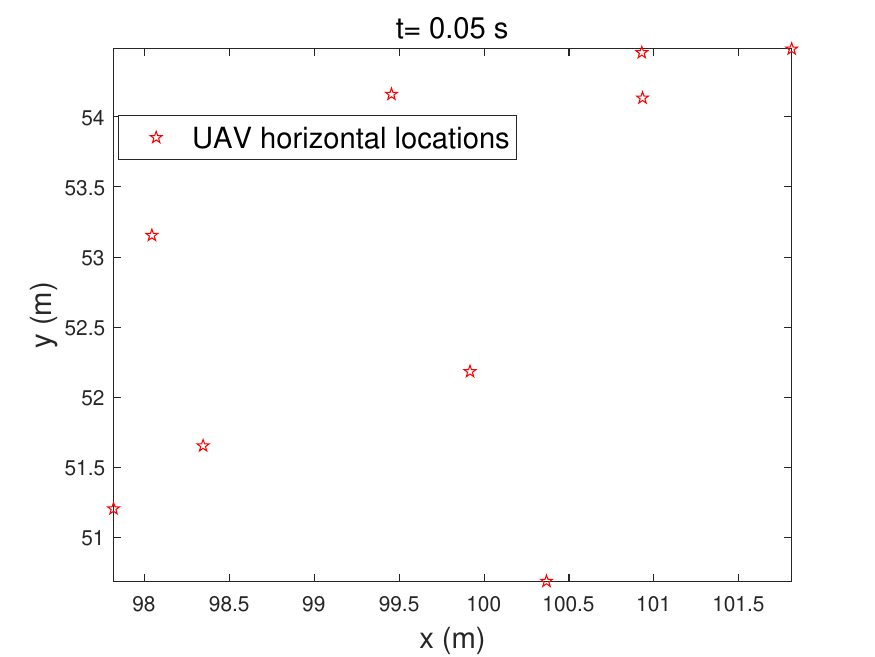}
       \caption{Locations, $t=0.05$ s.}
    \label{figsubber1}
  \end{subfigure}%
\begin{subfigure}{0.5\columnwidth}
  \centering
\includegraphics[width=\linewidth]{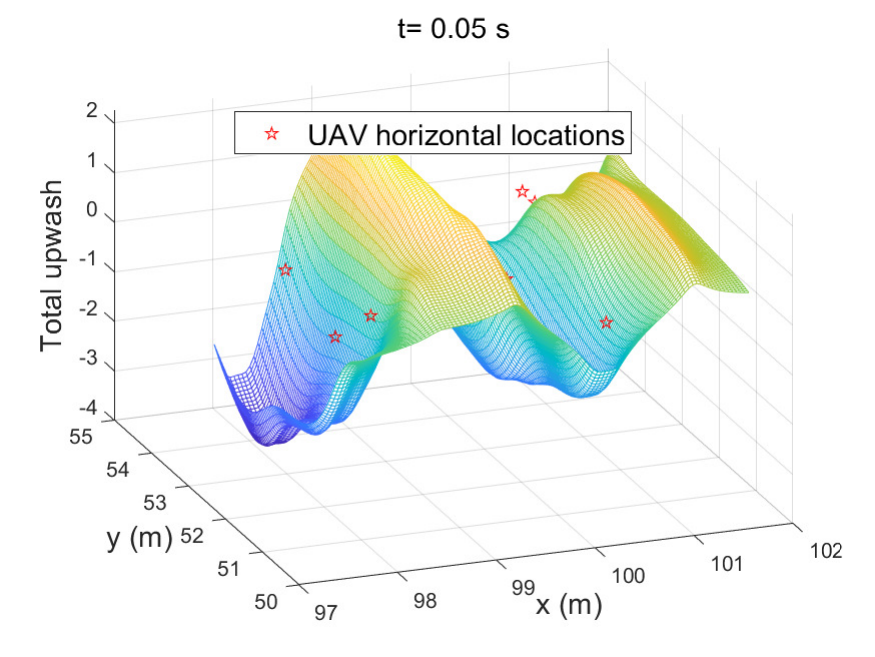}
    \caption{Total upwash, $t=0.05$ s.}
    \label{figsubcrb1}
  \end{subfigure}
\begin{subfigure}{0.5\columnwidth}
    \centering 
\includegraphics[width=\linewidth]{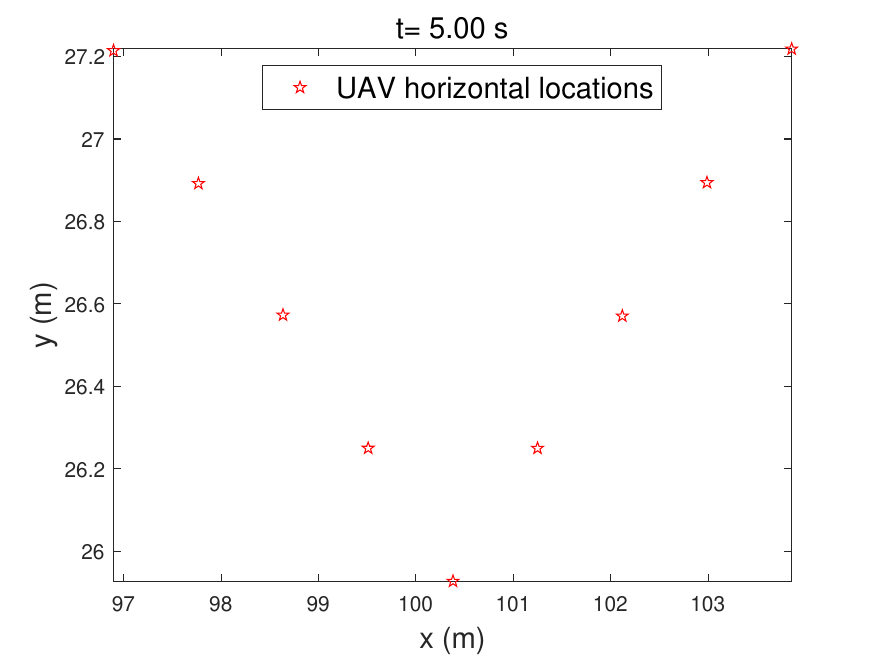}
    \caption{Locations, $t=5$ s. }
    \label{LOCA_5S}
  \end{subfigure}
  \begin{subfigure}{0.5\columnwidth}
    \centering
\includegraphics[width=\linewidth]{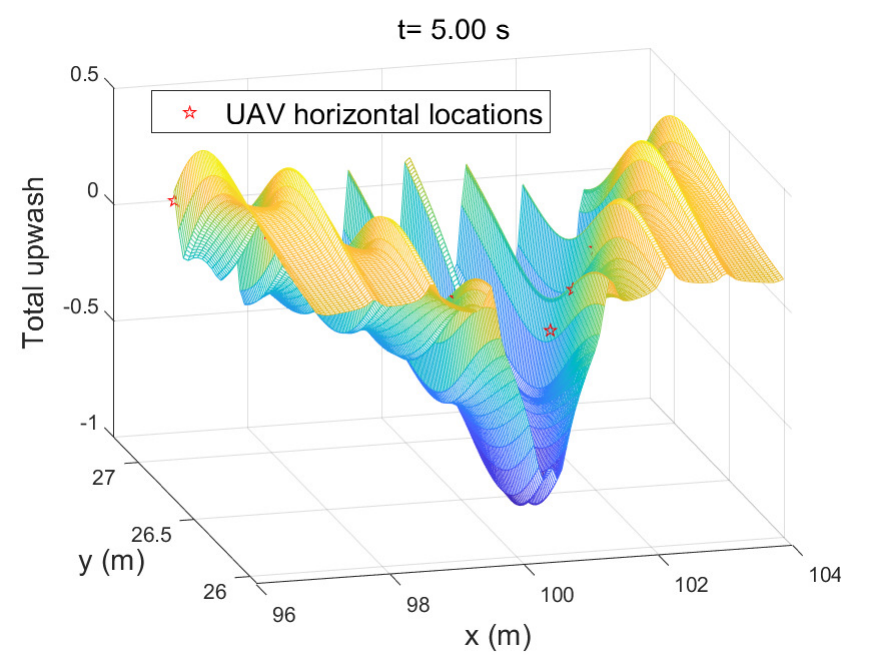}
\caption{Total upwash, $t=5$ s.} \label{upwash999}
  \end{subfigure}
\captionsetup{font=small,name=Fig,labelsep=period,justification=justified,singlelinecheck=false}
    \caption{The horizontal locations and the total upwash of formation 2 including $9$ UAVs in various time slots. }
    \label{formation2}
    \vspace{-0.5cm}
  \end{figure*}
We first evaluate the formation behavior of these two UAV formations in different time slots, as depicted in Fig. \ref{figsys} and Fig. \ref{formation2}, respectively. The horizontal locations of the first formation (with $19$ UAVs) are initialized around the center $[20 \ 50]^\mathrm{T}$ m. As observed, the upwash in $t=0.05$ s is tanglesome, and most UAVs suffer from downwash, increasing the power consumption to overcome the induced drag. As time passes, all the UAVs invoke Algorithm $1$ to adjust their locations by exchanging information with neighbors. Although two maximum upwash points may exist for the follower, the optimal destinations, where the UAVs enjoy the upwash generated by the remaining UAVs, can be uniquely determined by incorporating the variable $\Lambda_{m,k}$. It is observed that the formation finally converges to a `V' shape in $t=10$ s, which is consistent with the flight of bird flock immigration \cite{weimerskirch2001energy}, confirming the effectiveness of the proposed approach. By doing so, each UAV positions itself within the maximum upwash zone relative to its reference UAV. It is noted that the UAVs may not be able to achieve the global optimal positions due to the first-order Taylor expansion in \eqref{grad}. However, our proposed design can still guarantee effective exploitation of the upwash, reducing flight energy consumption. Similarly, in Fig. \ref{formation2}, we show the behavior of formation $2$ with initialized horizontal locations around the center $[100\ 50]^\mathrm{T}$ m, which also demonstrates a `V' shape when the formation becomes stable. We observe that formation $2$ converges no more than $5$s compared to formation $1$. This is as expected since only $9$ UAVs are involved within formation $2$, indicating a less information exchange process among neighbors. \revise{More interestingly, Fig. \ref{upwash19} and Fig. \ref{upwash999} demonstrate that the leader UAVs in both formations are consistently exposed to pronounced downwash effects, due to the absence of preceding UAVs capable of generating beneficial upwash. This aerodynamic disadvantage increases induced drag, thereby elevating energy consumption and reducing flight velocity relative to the trailing UAVs. To mitigate this imbalance, a dynamic leader rotation mechanism could be introduced, enabling UAVs to periodically exchange leadership roles, which is left to our future work.}

  \begin{figure}
  \centering
      \includegraphics[width=0.9\linewidth]{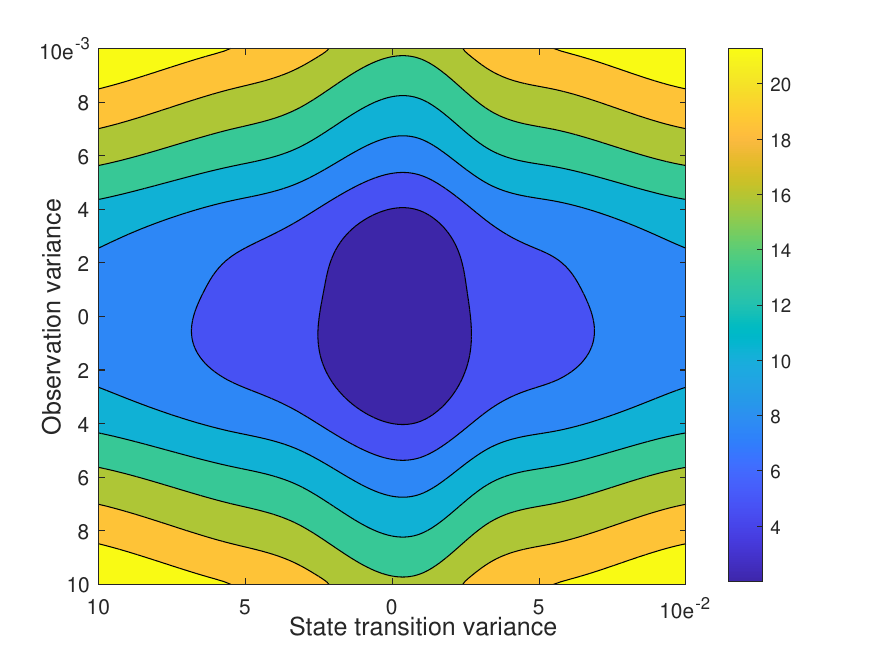}
      \caption{The LQR cost under various values of state transition noise variance $\mathbf{\Sigma}_{\mathrm{v}}$ and observation noise variance $\mathbf{\Sigma}_\omega$ with $P_{\max}=30$ dBm and $\Gamma=0 $ dBm.}
      \label{figobsnoisestatnois}
      \vspace{-0.5cm}
  \end{figure}
Next, we show the LQR cost with different observation noise variance and state transition noise variance with the maximum transmit power $P_{\max}=30$ dBm in Fig.
\ref{figobsnoisestatnois}. It can be seen that the LQR cost increases with both $\mathbf{\Sigma}_{\mathrm{v}}$ and $\mathbf{\Sigma}_\omega$, reflecting the compounding negative impact of increased state transition and observation noise on system performance.
The central region of Fig.
\ref{figobsnoisestatnois} demonstrates a combination of $\mathbf{\Sigma}_{\mathrm{v}}$ and $\mathbf{\Sigma}_\omega$ where the LQR cost is minimized. \revise{It is noteworthy that the LQR cost in Fig. \ref{figobsnoisestatnois} does not vanish even when both the observation noise covariance $\boldsymbol{\Sigma}_{\mathrm{v}}$ and the state transition noise covariance $\boldsymbol{\Sigma}_\omega$ approach zero. This phenomenon arises due to the intrinsic capacity limitation of the wireless communication channel. As a result, even in the absence of process and measurement noise, the controller cannot perfectly track or regulate the system state due to limited information precision and quantization effects, highlighting the fundamental trade-off between control performance and communication constraints.} Additionally, the LQR cost presents a high sensitivity to changes in 
$\mathbf{\Sigma}_{\mathrm{v}}$ and $\mathbf{\Sigma}_\omega$  in certain regions. For instance, as $\mathbf{\Sigma}_\omega$ increases beyond a certain threshold (e.g., $5\times10^{-3}$ ), the LQR cost rises significantly. Conversely, under relatively small variations, the regions have flatter contours.

    \begin{figure}
      \centering  \includegraphics[width=0.9\linewidth]{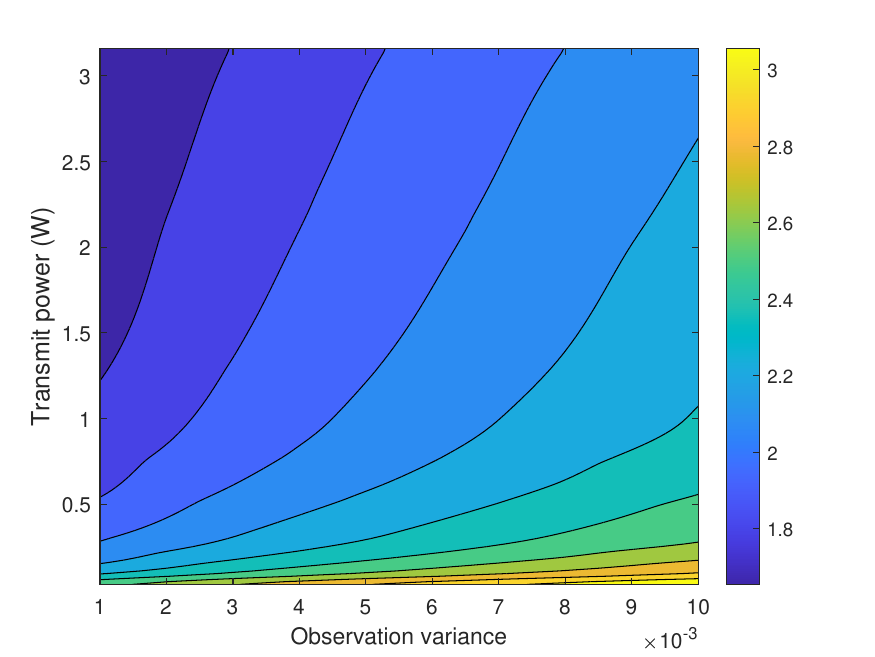}
      \caption{The LQR cost with different values of $\mathbf{\Sigma}_\omega$ and transmit powers.}
      \label{figpowobsnoe}

  \end{figure}
We further illustrate the LQR cost under different values of $\mathbf{\Sigma}_\omega$ and transmit powers in Fig. \ref{figpowobsnoe}. This demonstrates the critical trade-offs between transmit power and the LQR cost in the presence of observation noise, providing valuable insights for robust control systems in real-world environments. As expected, the LQR cost increases with higher observation noise variance, indicating the degradation in control performance when observations are unreliable. Furthermore, we can observe that the control performance is enhanced as the transmit power budget becomes generous. This phenomenon arises from the fact that a higher transmit power budget can achieve a higher channel capacity, enhancing the control performance in terms of LQR based on \eqref{eq:rate_cost_tradeoff-1}. In addition, the LQR would converge to $\bar{l}_{k,\min}$ if the transmit power is further improved.

 \begin{figure}
      \centering  \includegraphics[width=0.9\linewidth]{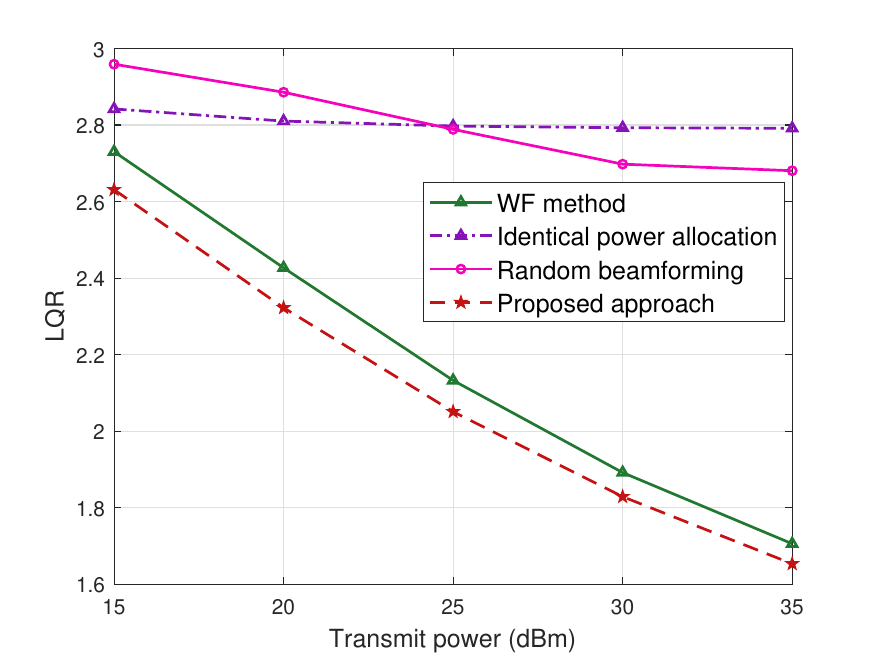}
      \caption{The LQR vs. transmit powers with various benchmark schemes.}
      \label{figlqrbenchmark}
  \end{figure}
In Fig. \ref{figlqrbenchmark}, we demonstrate the LQR cost versus different transmit powers with the sensing threshold $\Gamma=0$ dBm. In particular, we compare the control performance of our proposed design with three benchmarks, i.e., the water-filling (WF) method, identical power allocation, and random beamforming approach. As observed, the proposed approach outperforms other benchmarks across all transmit power levels due to the designed beamformer, highlighting its robustness and effectiveness in LQR cost reduction. Moreover, the WF method provides a viable alternative, which is because it can effectively improve the sum rate for the two leaders. However, the WF method cannot guarantee sufficient throughput for the weak user, compromising control fairness and causing performance degradation. Additionally, increasing power can only slightly impact the LQR of the identical power allocation scheme. This is due to the fact that the received power and interference are concurrently enhanced.
  
 \begin{figure}
      \centering  \includegraphics[width=0.9\linewidth]{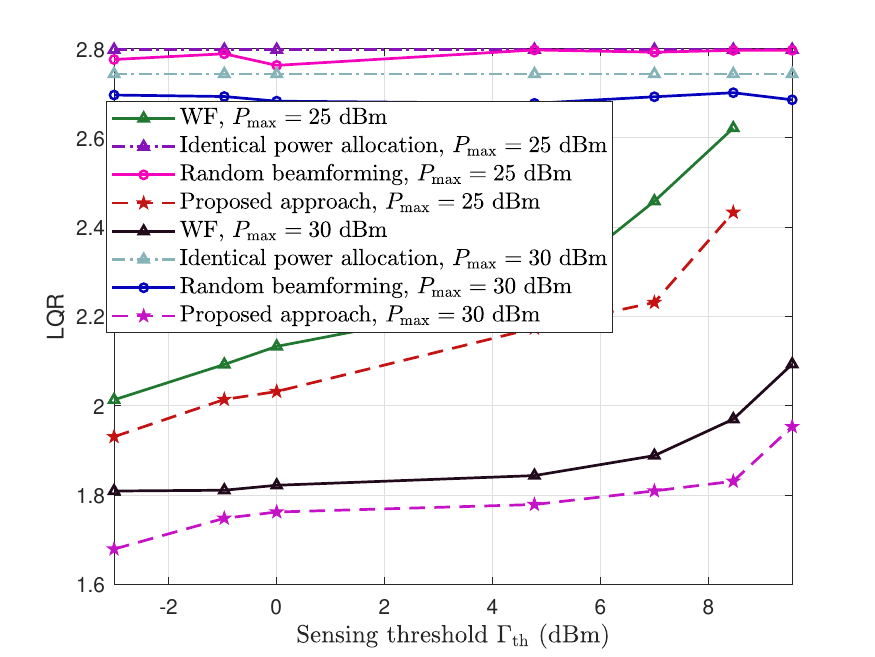}
      \caption{\revise{The LQR cost versus various sensing thresholds.}}
      \label{fig:sensingthreshold}
  \end{figure}
To further investigate the trade-offs between sensing accuracy and control performance, we present the LQR cost for different sensing thresholds in Fig. \ref{fig:sensingthreshold}. As shown, the system experiences a degradation in control performance as the required sensing threshold increases. This is because a higher sensing threshold \( \Gamma_\mathrm{th} \) imposes a more stringent sensing constraint, which in turn necessitates a greater allocation of power to meet the sensing requirements, thereby compromising the control performance. For instance, for the case of \( P_\mathrm{max} = 25 \) dBm, the beamforming design becomes infeasible when \( \Gamma_\mathrm{th} \) exceeds a certain threshold (e.g., \( \Gamma_\mathrm{th} = 9 \) dBm). In addition, the LQR costs associated with the random beamforming and identical power allocation schemes remain nearly constant, regardless of the variation in \( \Gamma_\mathrm{th} \), as they do not consider the sensing requirement.
\begin{figure}
    \centering
    \includegraphics[width=0.9\linewidth]{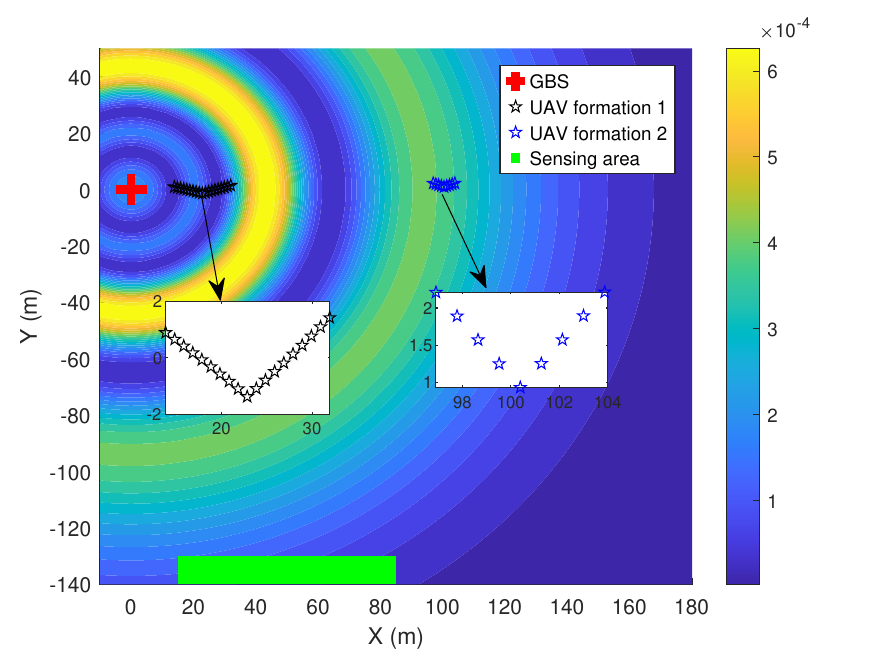}
    \caption{The sensing beam pattern gain at $t=10$ s with sensing thresholds $\Gamma_\mathrm{th}=0$ dBm.}
    \label{fig:enter-label}
\end{figure}

Fig. \ref{fig:enter-label} illustrates the sensing beam pattern gain at a specific time \( t = 10 \, \text{s} \) under the sensing threshold of \( \Gamma_{\text{th}} = 0 \, \text{dBm} \). It reveals that the sensing beam pattern gain is predominantly concentrated around UAV formation 1, as formation 2 is positioned farther from the central region. This highlights the impact of UAV placement on the spatial distribution of beam pattern gain. Furthermore,  while the sensing area does not receive the maximum illumination power at this specific time instant, it does not necessarily compromise the overall sensing performance. The reason is that the beamformer is designed to ensure sufficient sensing performance over the entire duration, rather than achieving peak performance in a single time slot.
\begin{figure}
    \centering
    \includegraphics[width=\linewidth]{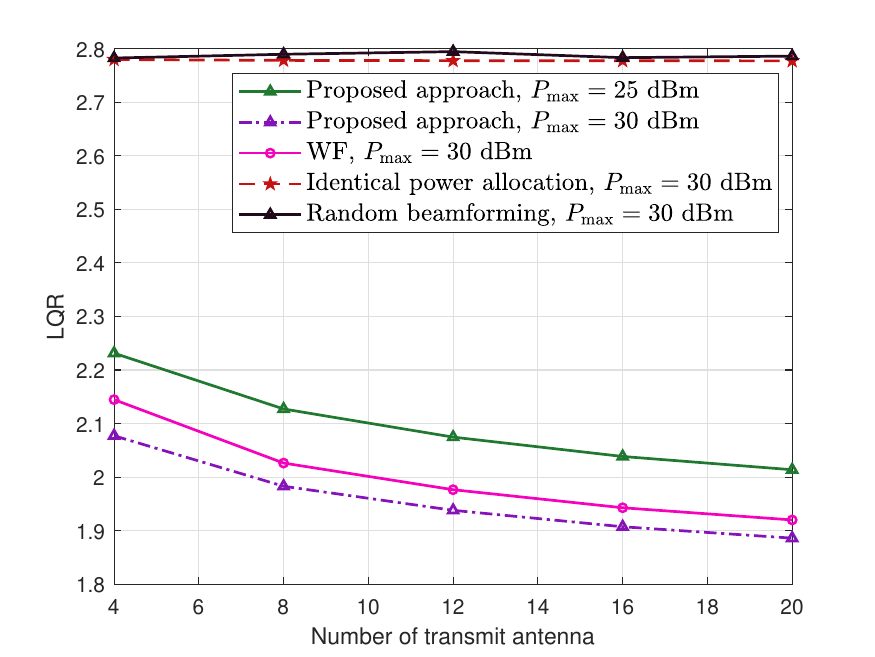}
    \caption{\revise{The LQR vs. transmit antenna numbers with various benchmark schemes.}}
    \label{fignum}
\end{figure}
\revise{Fig. \ref{fignum} demonstrates that increasing the number of antennas leads to a consistent reduction in the LQR cost across all schemes. This improvement stems from enhanced beamforming capability and spatial selectivity, which increase SINR and improve control signal reliability. The proposed approach outperforms all baselines, and the performance gain becomes more significant as the antenna array size grows. In contrast, schemes with fixed or random power allocation show limited sensitivity to antenna scaling due to their inability to exploit the spatial degrees of freedom effectively.}

\section{Conclusion} \label{seccon}
This paper investigated system design aimed at optimizing wireless control performance for energy-saving UAV formations tailored for dual-functional LAWNs.  By leveraging the aerodynamic upwash effect, we first developed a distributed energy-saving formation framework by capitalizing on the ATC diffusion LMS approach. Specifically, each UAV utilizes the LMS algorithm to update the local estimate and then refines the ultimate decisions by exchanging information with neighboring UAVs. To guarantee the control performance, we formulated an LQR minimization problem constrained by the available power budget and required sensing beam pattern gain toward the target area in low-altitude airspace. To address this non-convex optimization problem, we proposed an efficient iterative algorithm to obtain a near-optimal solution by employing the SCA and SDR techniques. Extensive simulation results showed that the `V' formation is likely to be the most effective energy-saving formation and demonstrated the superiority of our proposed method compared to benchmarks in terms of LQR cost.

   \bibliographystyle{IEEEtran}
\bibliography{citation}

\end{document}